\documentstyle[12pt,graphicx]{article}
\title{Approximate Analytic Spectra of Reionized CMB Anisotropies and Polarization
generated by Relic Gravitational Waves }
\author{T.Y. Xia  and  Y. Zhang \thanks{yzh@ustc.edu.cn} \\
                Key Laboratory of Galactic and Cosmological Research\\
                Astrophysics Center \\
        The University of Sciences and Technology of China\\
        Chinese Academy of Sciences\\
        Hefei, Anhui, China }
 \date{}

\begin{document}
\maketitle
\baselineskip=21truept

\newcommand{\be}{\begin{equation}}
\newcommand{\ee}{\end{equation}}
\newcommand{\ba}{\begin{eqnarray}}
\newcommand{\ea}{\end{eqnarray}}

\sf
\begin{center}
\Large  Abstract
\end{center}
\begin{quote}
 {\large
We present an approximate, analytical calculation of the reionized spectra
$C_l^{XX}$ of cosmic microwave background radiation (CMB)
anisotropies and polarizations generated by relic gravitational
waves (RGWs).
Three simple models of reionization are explored,
whose visibility functions are fitted by gaussian type of functions
as approximations.
We have derived the analytical polarization
$\beta_l$ and temperature anisotropies $\alpha_l$,
both consisting of two terms proportional to RGWs
at the decoupling and at the reionization as well.
The explicit dependence of  $\beta_l$ and $\alpha_l$
upon the reionization time $\eta_r$,
the duration $\Delta\eta_r$,
and the optical depth $\kappa_r$  are demonstrated.
Moreover, $\beta_l$ and $\alpha_l$ contain $\kappa_r$
in different coefficients,
and the polarization spectra $C_l^{EE}$ are $C_l^{BB}$ are more sensitive
probes of reionization than $C_l^{TT}$.
These results facilitate examination of the reionization effects,
in particular,
the degeneracies of $\kappa_r$ with the normalization
amplitude and with the initial spectral index of RGWs.
It is also found that
reionization also causes a
$\kappa_r$-dependent shift  $\Delta l\sim 20$ of the zero multipole
$l_0$ of $C_l^{TE}$,
an effect that should be included in order to detect the traces of RGWs.
Compared with numerical results,
the analytical $C_l^{XX}$ as approximation
have the limitation.
For the primary peaks in the range $l\simeq (30, 600)$,
the error is $\le 3\%$ in three models.
In the range $l < 20$ for the reionization bumps,
the error is $\le 15\%$ for $C_l^{EE}$ and  $C_l^{BB}$
in the two extended  reionization models,
and $C_l^{TT}$ and  $C_l^{TE}$ have much larger departures for $l<10$.
The bumps in the sudden reionization model are too low.
}

\end{quote}


Key words:

\newpage
\baselineskip=19truept

\begin{center}
{\em\Large 1. Introduction}
\end{center}

Reionization is a very important cosmological process, which might
be, to a large extent, determined by the first luminous objects
formed in the early universe, either star-forming galaxies or
active galactic nuclei.
Our knowledge of the cosmic structure
formation of the universe would be incomplete without a reliable
account of reionization history, the details of which is still not
understood yet. During the evolution history of  CMB, the
reionization taking place around the redshift
$z= (20 \sim  6)$ is a
major process in shaping the profiles of CMB spectra on large
scales, only secondary to the decoupling around $z\sim 1100$.
Reionization leaves observable prints on CMB
\cite{Spergel03,Peiris,spergel,Page,komatsu,nolta,hinshaw,dunkley}
through the interaction between the CMB photons and the reionized
free electrons.
In particular, the spectra of CMB anisotropies
and polarizations on large angular scales contain the distinguished
signatures of reionization. Thereby, complementary to the
constraints on the late stage of reionization $z\simeq 6$ from
observations of the most distant quasars absorption lines, etc, CMB
provides a unique probe for the early stage of reionization.
On the other hand, in order to interpret the observed
spectra of CMB anisotropies and polarizations within the standard
model, the reionization-induced modifications have to be taken into
account properly.
As is known, the reionization parameters could be
entangled with the cosmological parameters, thus biasing our
interpretation of CMB, and of reionization as well
\cite{MZaldarriaga,ZSS,ZS,Venkatesan1,Venkatesan2,NgNg}. In this
regards, analytic studies can improve our  understanding of CMB and
reionization, even though the comparisons with the observed data
need more accurate numerical calculations, such as cmbfast and CAMB
\cite{Seljak,Lewis}.

Two kinds of perturbations of the spacetime metric,
i.e., density perturbation \cite{Sasaki,Muhhanov,Hu95a}
and relic gravitational waves (RGWs)
\cite{Muhhanov,Basko,Polnarev,Crittenden},
will effectively influence the CMB
through the Sachs-Wolfe term \cite{Sachs} in
the Boltzmann equation for photons.
Although the contribution by density perturbation is dominant,
RGWs give rise to a magnetic type of CMB polarizations,
providing a distinguished channel
to directly detect RGWs of very long wavelength
\cite{MZaldarriaga,ZS,Seljak3,Kamionkowski}.
Moreover, RGWs have substantial contributions
to large angular scales part of CMB spectra,
where the impact of reionization is also dominant and
cause bumps in the CMB polarization spectra for  $l<10$.
Thus,
in order to study reionization through the CMB,
one has to take into account of the contribution of RGWs,
or, vice versa.

The analyses have been made towards
CMB anisotropies and polarizations generated by RGWs
\cite{Basko,Polnarev,Keating1,Keating,
Kamionkowski,Kosowsky,Zaldarriaga,Grishchuk,Zhao09}.
In particular, by an approximate treatment of the time integration
over the decoupling process during the recombination,
Refs.\cite{Pritchard04,Zhao06}
have derived the analytic expressions of the
 CMB polarization  spectra, $C_l^{EE}$ and $C_l^{BB}$.
Recently, extending the previous works,
we have improved the time integration by a better approximation,
and obtained the analytical expressions of
all the four spectra, including $C_l^{TT}$ and $C_l^{TE}$ \cite{Xia},
which agree fairly with the numerical results
up to a broader range of multipole moment $l<600$.
In that work the damping on RGWs due to neutrino free-streaming (NFS)
has been included \cite{Weinberg,Dicus,Watanabe,Miao07},
and its effects on the cross spectrum $C_l^{TE}$ have been demonstrated
in details.
In these analytical calculations,
the reionization process has not bee included,
which will be addressed in this paper.
For the purpose of calculating the reionized CMB spectra $C_l^{XX}$,
the reionization can be treated similarly to the decoupling,
if the visibility functions for both processes are given.
While the decoupling and its visibility function $V_d(\eta)$
effectively distributed around $z\sim 1100$ have been better studied,
the reionization is currently less understood,
and is commonly modeled by
its ionization fraction $X_e(\eta)$ as a function of time.
We shall examine three possible simple reionization models
with explicit $X_e(\eta)$,
which, for a given value of the optical depth $\kappa_r$,
can be converted into
its corresponding visibility function $V_r(\eta)$
effectively distributed around $z\sim 11$.
The functions $V_d(\eta)$ and $V_r(\eta)$
are separately distributed, not overlapping,
each of them can respectively
be approximated by Gaussian type of functions,
which are specified by their location, height, and width.
In parallel,
we will carry out, with approximation,
the time integrations of Boltzmann's equation
for the decoupling and reionization processes.
The modes $\alpha_l$ and $\beta_l$,
respectively, for
CMB temperature anisotropies and polarization,
are obtained as analytical expressions.
Each mode explicitly consists of two separated parts,
one from the decoupling, and another from  the reionization.
Moreover, the optical depth $\kappa_r$
appears as the coefficients in  $\alpha_l$ and $\beta_l$
in different combinations,
and probabilistic interpretations are given.
Besides reionization and  decoupling,
the result contains also other cosmological parameters
for  inflation that are contained in RGWs.
Thus analytic studies on the reionization effects
will be  facilitated.

In Section 2
we review briefly the result of RGWs spectrum $h(\nu,\eta)$
that will  be used as the source for
CMB anisotropies and polarization.
In Section 3
three models of homogeneous reionization  $X_e(\eta)$
are presented:
one sudden  and two extended.
For each model the visibility function $V_r(\eta)$
and the optical depth function $\kappa_r(\eta)$ are presented.
In Section 4,
by  approximately carrying out the time integrations,
the analytical expressions of  $\alpha_l$
and $\beta_l$ are obtained.
The resulting spectra  $C_l^{XX}$ are demonstrated.
In Section 5,
detailed analyses are made towards the reionization effects
upon  $C_l^{XX}$,
three models are compared,
and, in particular,
examinations are made on the degeneracies of $\kappa_r$
with the normalization amplitude $A$
and the initial spectral index $\beta_{inf}$ of RGWs
produced during inflation.
The effect of reionization on the zero multipole analysis
is addressed.
The  conclusion is given in Section 6.
We use the unit in which $c=\hbar =k_B= 1$ in this paper.

\begin{center}
{\em\Large 2. RGWs Spectrum}
\end{center}

The expansion of a spatially flat Universe
can be described by the spatially flat
($\Omega_\Lambda +\Omega_m+\Omega_r=1$) Robertson-Walker spacetime
with a metric
\be
ds^2=a^2(\eta)[-d\eta^2+(\delta_{ij}+h_{ij})dx^idx^j],
\ee
where $a(\eta)$ is the scale factor,
$\eta$ is  the conformal time,
and $h_{ij}$ is the gravitational waves,
taken to be  traceless and transverse (TT gauge)
$h^i_{\,\,i}=0$, and $h_{ij,j}=0$.
By the Fourier decomposition
\be \label{hij}
h_{ij} (\eta,{\bf x})=
\sum_{\sigma}\int\frac{d^3k}{(2\pi)^3}
\epsilon^{\sigma}_{ij}h_{\bf k}^{(\sigma)}(\eta)
e^{i\bf{k}\cdot{x}}
\ee
for each mode $\bf k$ and each polarization $\sigma =( +,\times)$,
the wave equation takes the form
\be \label{heq}
\ddot{h}_k+2\frac{\dot{a}}{a}\dot{h}_k+k^2h_k=0,
\ee
where the polarization index $\sigma$
has been skipped for simplicity,
and the subindex $\bf k$ can be replace by  $k$
since the perturbations are assumed to be isotropic.
The analytic solution of Eq.(\ref{heq}) has been given
for the expanding universe with the consecutive stages:
inflationary, reheating,
radiation-dominant, matter-dominant,
and accelerating, respectively
in Refs.\cite{Miao07,zhang06b,wangzhang08}.
In our convention,
\be \label{m}
a(\eta)=a_m(\eta-\eta_m)^2,\,\,\,\,\eta_2 \leq \eta\leq \eta_E,
\ee
for the matter-dominant stage, and
\be \label{accel}
a(\eta)=l_H|\eta-\eta_a|^{-\gamma},\,\,\,\,\eta_E \leq \eta\leq \eta_0,
\ee
for accelerating stage up to the present time $\eta_0$,
where  $\gamma \simeq 1.044$ for $\Omega_{\Lambda}=0.75 $,
and $l_H=\gamma/H_0$, $H_0$ is the Hubble constant.
The normalization of $a(\eta)$ is chosen to be $|\eta_0-\eta_a|=\eta_a-\eta_0=1$,
where we have taken $\eta_0=3.11$ to be the present time.
Then, once the ratio $\Omega_\Lambda/\Omega_m$ is specified,
all the parameters will fixed:
$a_m=l_H\frac{\gamma^2}{4}\zeta_E^{-(1+2/\gamma)}$,
$\eta_E=\eta_a-\zeta_E^{1/\gamma}$,
$\eta_m=\eta_E-\frac{2}{\gamma}\zeta_E^{1/\gamma}$
with $\zeta_E\equiv (\Omega_\Lambda/\Omega_m)^{1/3}$.
The details have been explicitly
demonstrated in our previous study of RGWs  \cite{Miao07,zhang06b}.

When the NFS is included,
a process occurred
from a temperature $T \simeq 2$ MeV during the radiation stage
 up to the beginning of the matter domination,
the analytic solution  $h_k(\eta)$ has been given \cite{Xia,Miao07}.
The NFS
causes a damping of the amplitude of RGWs by $\sim 20\%$
in the frequency range $(10^{-17}, 10^{-10})$ Hz,
leaving observable signatures
on the second and third peaks of CMB anisotropies and polarization.
So the RGWs damped by NFS will be used as a source in our calculation.
As for other physical processes,
such as the QCD transition and the $e^{\pm}$ annihilation
in the radiation stage \cite{Watanabe,wangzhang08,schwarz},
they only cause minor modifications
of RWGs on the small scales $\nu>10^{-12}$ Hz,
not being observable in the present large-scale CMB spectra,
and will not be considered here.

The solution $h_k(\eta)$ depends on
the initial condition during the inflation stage.
We choose the initial spectrum of RGWs
at the time $\eta_i$ of the horizon-crossing
\cite{Xia,Miao07,zhang06b,Grishchuk01}
\be \label{initialspectrum}
h(\nu, \eta_i) =\frac{2k^{3/2}}{\pi}|h_k(\eta_i)|
=A(\frac{k}{k_H})^{2+\beta_{inf}},
\ee
where  $k_H \simeq 2\pi$ is the comoving wavenumber
corresponding to the Hubble radius,
 $A$ is a $k$-independent constant
to be normalized by the present observed CMB anisotropies in practice,
and the spectral index $\beta_{inf}$ is
a parameter depending on inflationary models.
The special case of $\beta_{inf}=-2$ is
the de Sitter expansion of inflation.
If the inflationary expansion is driven by a scalar field,
then the index $\beta_{inf}$ is related to
the so-called slow-roll parameters,
$\eta$ and $\epsilon$ \cite{Liddle},
as $\beta_{inf}=-2+(\eta-3\epsilon)$.
$\beta_{inf}$ is related to the spectral index $n_S$
of primordial scalar perturbations as
$n_S=2\beta_{inf} +5$.
In literature,  the RGWs spectrum is also written
in the following form \cite{Spergel03} \cite{Peiris} \cite{Verde}
\be \label{Delta}
\Delta^2_h(k)=A_T(\frac{k}{k_0})^{n_T}= \frac{1}{8}h^2(\nu, \eta_i),
\ee
where the tensor spectrum index
$n_T=2(\beta_{inf}+2)\sim 0$ without the running index,
$k_0$ is some pivot wavenumber,
taken as $k_0= 0.002$ Mpc$^{-1}$ in our calculation,
and the tensor spectrum amplitude
$A_T=2.95\times10^{-9}A(k_0)\, r$,
where $A(k_0)$ is the scalar power spectrum amplitude
that can be determined by the WMAP observations
\cite{Spergel03,spergel,Page},
and we take $A(k_0)\sim 0.8$ accordingly.
The tensor/scalar ratio $r$ is model-dependent,
and  frequency-dependent \cite{Zhao06,baskaran}.
Recently, the 5-year WMAP data improves the upper limit to
$r<0.43$ ($95\%$ CL) \cite{dunkley}, and combined with BAO and SN
gives $r<0.2$ ($95\%$ CL) \cite{komatsu} \cite{hinshaw}.
In our treatment, for simplicity,
 $r\simeq 0.37$ is only taken as a
constant parameter for normalization of RGWs,
except otherwise mentioned.

The resulting functions $h_k(\eta)$  and $\dot{h}_k(\eta)$
serve as the tensorial source to CMB anisotropies and polarization.
Without reionization,
only RGWs $h_k(\eta_d)$  and $\dot{h}_k(\eta_d)$
at the decoupling time $\eta_d$
are relevant, contributing to the primary CMB spectra.
When reionization comes,
$h_k(\eta_r)$ and $\dot{h}_k(\eta_r)$
at the reionization $\eta_r$
contribute too,
mainly contributing to the very large angular reionization bumps of CMB spectra.
In Fig.\ref{fig1},
$h_k(\eta_d)$  and $\dot{h}_k(\eta_d)$,
and $h_k(\eta_r)$ and $\dot{h}_k(\eta_r)$
are plotted.
The right panel of Fig. \ref{fig1} shows that,
$\dot{h}(\eta_d)$ has the greatest amplitude around $k\sim 25$,
forming a deep trough,
whereas $\dot{h}(\eta_r)$ has the greatest amplitude
around $k\sim 2$, forming a deep trough.
The left panel shows that both $h_k(\eta_d)$ and $h_k(\eta_r)$
have similar slope for small $k$.
As we will see,
these features of RGWs at $\eta_d$ and at $\eta_r$
are responsible  for the profiles  of CMB spectra  $C_l^{XX}$.

\begin{figure}
\caption{  \label{fig1}
The RGWs $h_k(\eta_d)$ and $\dot{h}_k(\eta_d)$
at the decoupling
and $h_k(\eta_r)$ and $\dot{h}_k(\eta_r)$
at the reionization.
}
\end{figure}

\begin{center}
{\em\Large 3. Visibility Function}
\end{center}

In Basko and Ponarev's method,
the Boltzmann equation of the photon gas for the $k$-mode
is written as a set of two coupled differential equations
\cite{Basko,Polnarev}
\be \label{eqxi}
\dot{\xi}_k+[ik\mu +q]\xi_k= \dot{h}_k,
\ee
\be \label{eqbeta}
\dot{\beta_k}+[ik\mu +q]\beta_k=qG_k.
\ee
where  $\beta_k$ is the linear polarization
contributed only by linearly polarized CMB photons,
\be \label{alpha}
\alpha_k \equiv\xi_k-\beta_k
\ee
is the anisotropy of radiation intensity contributed
by both unpolarized (natural light) and polarized CMB photons,
$\mu= \cos\theta$,
 $q$ is the differential optical depth,
and
\be\label{G}
G_k(\eta)=\frac{3}{16}
\int^1_{-1}d \mu'[(1+\mu'^2)^2\beta_k-\frac{1}{2}(1-\mu'^2)^2\xi_k].
\ee
In the following,
we omit the subscript $k$ for simplicity of notation.
The formal solutions of Eqs.(\ref{eqxi}) and (\ref{eqbeta})
at any time $\eta$ can be written as the following time integrations
\cite{Zhao06,Xia}:
\be \label{xi}
\xi(\eta)=\int^{\eta}_0 \dot{h}(\eta')
e^{-\kappa(\eta,\eta')}e^{ik\mu(\eta'-\eta)}d \eta',
\ee
\be \label{beta}
\beta(\eta)=\int^{\mu}_0 G(\eta')q(\eta')
e^{-\kappa(\eta,\eta')}e^{ik\mu(\eta'-\eta)}d \eta',
\ee
where
\be
\kappa(\eta',\eta)\equiv \int_\eta^{\eta'} qd\eta
=\kappa(\eta)-\kappa(\eta')
\ee
with the optical depth given by
\be \label{kappa-definition}
\kappa(\eta)\equiv\kappa(\eta_0,\eta)
=\int_{\eta}^{\eta_0} q(\eta')d\eta'
\ee
from the present time $\eta_0$ back to an earlier  time $\eta$,
such that
\be
q(\eta)= -\frac{d \kappa(\eta)}{d\eta}.
\ee
The CMB anisotropies and polarization
are usually expressed in terms of their Legendre components
\be\label{xill}
\xi_l(\eta) = \frac{1}{2}\int_{-1}^1 \,d\mu\,
\xi(\eta,\mu)P_l(\mu),
\ee
\be\label{betall}
\beta_l(\eta) =\frac{1}{2}\int_{-1}^1\, d\mu\,
\beta(\eta,\mu)P_l(\mu),
\ee
where $P_l$ is the Legendre function.
By the  expansion formula
\be \label{jl}
 e^{ix\mu}
      =\sum_{l=0 }^{\infty} (2l+1)i^lj_l(x)P_l(\mu)
\ee
and the ortho-normal relation for the Legendre functions,
the components at the present time $\eta_0$
are given by the following
\be \label{xigeneral}
\xi_l(\eta_0)=
i^l \int^{\eta_0}_{0}
    e^{-\kappa(\eta)}\dot{h}(\eta)j_l(k(\eta-\eta_0)) d\eta,
\ee
\be \label{betae}
\beta_l(\eta_0) =
  i^l \int_0^{\eta_0} G(\eta)V(\eta)\, j_l(k(\eta-\eta_0))d\eta,
\ee
 where
\be \label{visibilitydef}
V(\eta)=q(\eta)e^{-\kappa(\eta)}
\ee
is the visibility function.
As one sees, to analytically carry out the integrations
in Eqs.(\ref{xigeneral}) and (\ref{betae}),
one needs the explicit expression of $e^{-\kappa(\eta)}$ and $V(\eta)$,
which are determined by the whole history of ionization.
In the following we will give approximate formula of both
functions.

$V(\eta)$ has the meaning of the probability that
a CMB photon reaching us today
was last scattered by free electrons at the time $\eta$.
Without the reionization,
$V(\eta)$ would have only one sharp peak around  $z\sim 1100$
for the decoupling, and satisfies
the normalization condition
\be \label{normv}
\int_0^{\eta_0} V(\eta)d \eta =1.
\ee
When the reionization is included,
$V(\eta)$  will have, around $z\sim 11$, another peak.
If the universe was reionized twice,
say at $z\sim 6$ and $z\sim 16$,
\cite{Cen,Giannantonio,Colombo2},
$V(\eta)$ would  have double peaks
for reionization.
We  consider only the case of a single reionization in this paper.
Then, as a function of $\eta$,
$V(\eta)$ is mainly distributed around decoupling and reionization,
and is effectively vanishing
in the region far away from the peaks,
as shown in the Panel (d) in Fig.\ref{fig6}.
Thus the time integration of Eq.(\ref{normv})
can be practically split into two parts
\be \label{unitary}
\int^{\eta_{split}}_{0}V_d(\eta) d\eta+
\int^{\eta_0}_{\eta_{split}}V_r(\eta) d\eta =1,
\ee
where $V_d(\eta)$ and $V_r(\eta)$
are the  portions of $V(\eta)$
for decoupling and reionization, respectively,
and $\eta_{split} $ is some point
between decoupling and reionization with $V(\eta_{split})\simeq 0$.
In calculation we can take, say,  $\eta_{split}= 0.297$
corresponding to a redshift $z\simeq 100$.
In Eq.(\ref{unitary}), $\int^{\eta_{split}}_{0}V_d(\eta) d\eta$
is the area covered under the curve of $V_d(\eta)$,
and stands for the probability that a photon
was last scattered during the decoupling.
Similarly, $\int^{\eta_0}_{\eta_{split}}V_r(\eta) d\eta$
is the probability that a photon
was last re-scattered during the reionization,
i.e., the amount of CMB photons out of the total
that are rescattered.
According to Eq.(\ref{unitary}),
their sum is constrained to be unity.
This has a physical interpretation:
more CMB photons are last scattered around $\sim \eta_r$,
less will be last scattered around $\sim \eta_d$.
During reionization
the intrinsic anisotropies
of this portion of CMB photons were washed out,
and new polarizations were generated on large angular scales.
As we will see,
$\int^{\eta_0}_{\eta_{split}}V_r(\eta) d\eta$ depends essentially
on the optical depth up to the reionization.

\begin{figure}
\caption{
\label{fig2}
The visibility function $V_d(\eta)$ for the
decoupling around $z\sim 1100$.
Both the analytic and the fitting by two
half-gaussian functions are shown. }
\end{figure}

Now let us specify the visibility functions
$V_d(\eta)$ and $V_r(\eta)$.
First the decoupling process is better understood,
whose $V_d(\eta)$ has been given explicitly,
which depends the baryon fraction $\Omega_B$
 \cite{Hu95a,peebles68, jones-wise}.
As a function of time,
the profile of $V_d(\eta)$ itself looks like
a sharp peak around the decoupling $z\sim 1100$.
Thus, when it appears as a factor of the integrand in
the time integration (\ref{betae}) for
the polarization $\beta_l(\eta_0)$,
it actually plays a filtering role:
only the narrow time range around the decoupling
contributes substantially to the integral of Eq.(\ref{betae}).
To facilitate analytic calculations of CMB polarization,
$V_d(\eta)$ has been approximated by the following two pieces of
half gaussian function \cite{Zhao06,Xia}
\be\label{halfgaussian1}
V_d(\eta)=
  \left\{
\begin{array}{ll}
V(\eta_d) \exp\left(-\frac{(\eta-\eta_d)^2}{2
\Delta\eta_{d1}^2}\right),        ~~~(\eta\leq\eta_d) ,   \\
V(\eta_d) \exp\left(-\frac{(\eta-\eta_d)^2}{2
\Delta\eta_{d2}^2}\right),         ~~~(\eta>\eta_d),
\end{array}
     \right.
\ee
where  $\eta_d$ is the decoupling time,
which is taken $\eta_d = 0.0707$ corresponding to
a redshift $z_d=1100$,
$\Delta\eta_{d1}=0.00639$, $\Delta\eta_{d2}=0.0117$, and
$(\Delta\eta_{d1}+\Delta\eta_{d2})/2=\Delta\eta_{d}=0.00905$
 is the thickness of the decoupling.
Eq.(\ref{halfgaussian1}) improves
a single gaussian function  \cite{Pritchard04}
by $\sim 10\%$ in accuracy
and at the same time allows an analytic
treatment of the CMB polarization spectrum.
We have checked that the errors between
Eq.(\ref{halfgaussian1}) and the numerical formulae given in
 \cite{Hu95a,jones-wise} is very small, $\le  3.9\%$ in the whole range.
The coefficient $V(\eta_d) $, as the height of $V_d(\eta)$,
also depends on the reionization
through the normalization in Eq.(\ref{unitary}).
The analytic $V_d(\eta)$ with $\Omega_B=0.046$
and its fitting are shown in Fig.\ref{fig2}.

Next,
understanding of the reionization as a physical process
is still underway,
and various tentative models have been proposed for it.
Spatially, the  reionization might have occurred
inhomogeneously \cite{Benson,Liu,Santos,JunZhang,Hu2007},
resulting in modifications on
the small angular scales part of CMB spectra.
Models of double reionization \cite{Cen,Giannantonio},
or its variants, such as peak-like reionization \cite{Naselsky},
have also been proposed.
In the following,
we will work with three simple homogeneous models,
whose ionization fraction $X_e(\eta)$ are explicitly given.
\begin{figure}
\caption{\label{fig3}
The three models of  reionization
with a fixed optical depth $\kappa_r=0.084$.
For each $X_e(\eta)$
given in Eqs.(\ref{sudden}), (\ref{lineargradual}),
and (\ref{Xeetalinear}),
the functions $q_r(\eta)$, $\kappa_r(\eta)$, and
$V_r(\eta)$ are calculated according to
the formulae in
Eqs.(\ref{q}), (\ref{kappar}), and (\ref{Vr}),  respectively.
}
\end{figure}
One is  the sudden reionization model
with
\ba \label{sudden}
X_e(\eta)=
  \left\{
\begin{array}{ll}
0 , &  \,\,\,  {\rm for } \,\,\,  \eta <\eta_{r}, \\
1 , &  \,\,\,  {\rm for } \,\,\,  \eta \geq \eta_{r} ,
\end{array}
     \right.
\ea
where $\eta_{r}$ is the reionization time.
For concreteness of illustration,
in our calculation we take $\eta_{r}=0.915$,
corresponding to the redshift $z_{r}=11$.
This is the simplest model often used in the literature.
But there are accumulating evidence
that the reionization is an  extended  process,
stretching  from $z\simeq 6$ up to $z\sim 11$,
even up to as early as $z\sim 20$ \cite{dunkley,Furlanetto,Hui}.
For instance,
studies of Ly$\alpha$ Gunn-Peterson absorption \cite{fan06}
indicate a rapid increase in the ionized fraction of
the intergalactic medium
at a redshift lower than $z_{r}\simeq 6$.
On the other hand,
the WMAP observations of CMB found a much earlier reionization,
$z_r= 17\pm 5$  by  WMAP 1-yr \cite{Spergel03},
$z_{r} = 10.9^{+2.7}_{-2.3}$ by  WMAP 3-yr \cite{spergel},
$z_{r} = 11.0\pm 1.4$ (68\% CL) by  WMAP 5-yr \cite{dunkley},
and $z_{r} = 10.8\pm 1.4$  by  WMAP 5-yr combined with SN and BAO
 \cite{hinshaw} \cite{komatsu}.
One extended reionization model is the $\eta$-linear  reionization with
\ba \label{lineargradual}
X_e(\eta)=
  \left\{
\begin{array}{ll}
0 , &  \,\,\,  {\rm for } \,\,\,  \eta <\eta_{r1} \\
\frac{ \eta-\eta_{r1} }{ \eta_{r2}-\eta_{r1} }, &  \,\,\,  {\rm for }
         \,\,\,  \eta_{r1}<\eta < \eta_{r2}, \\
1 , & \,\,\, {\rm for }\,\,\,  \eta > \eta_{r2} .
\end{array}
     \right.
\ea
where $\eta_{r1}$ and $\eta_{r2}$
are the beginning and end of reionization.
For instance, one can take $\eta_{r1}=0.685$ and $\eta_{r2}=1.20712$,
corresponding to $z_{r1}=20$ and $z_{r2}=6$, respectively.
This model is closer to the result
of WMAP 5-yr fitted by the two step reionization \cite{dunkley}.
Another extended reionization  model
is the $z$-linear model with \cite{Hu2007}:
\ba \label{Xeetalinear}
X_e(z)=
  \left\{
\begin{array}{ll}
0 , &  \,\,\,  {\rm for } \,\,\,  z > z_{r1} \\
 1-\frac{z -z_{r2}}{z_{r1}-z_{r2}}, &  \,\,\,  {\rm for }
         \,\,\,  z_{r1} >z> z_{r2}, \\
1 , & \,\,\, {\rm for }\,\,\, z \leq z_{r2} .
\end{array}
     \right.
\ea
For $z_{r1}=20$ and $z_{r2} =6$, one has $X_e(z)= 1-( z-6)/14$.
The ionization fraction $X_e (\eta)$ for these three reionization models
are comparatively shown in Fig.\ref{fig3}.

Given  $X_e(\eta)$ in the above three  models,
the differential optical depth for reionization
can be directly calculated by the formula \cite{Hu95a,Hu2007,peebles}:
\be
\label{q}
q_r(\eta)
=C_c \frac{a(\eta_0)^3}
{a(\eta)^2}X_e(\eta),
\ee
where the constant $C_c=(1-\frac{Y_P}{2})
          \frac{\Omega_b \rho_c \sigma_T}{m_p}$, $Y_p\simeq 0.23$
is the primordial helium fraction,
$\sigma_T$ is the cross section of Thompson scattering,
$m_p$ is the mass of a proton.
For $\Omega_b =  0.045$, $C_c\simeq 0.142 \times 10^{-28}$ m$^{-1}$.
Since the value of $Y_p$ from observations
has considerable large error bars \cite{Trotta},
in our treatment $C_c$ is allowed to vary slightly
around this value.
From Eq.(\ref{kappa-definition}) follows
the optical depth for reionization as an integration
\be\label{kappar}
\kappa_r(\eta)=\int_{\eta}^{\eta_0} q_r(\eta')d\eta',
\ee
and, from Eq.(\ref{visibilitydef}) follows
the visibility function for the reionization,
\be \label{Vr}
V_r(\eta) =q_r(\eta) e^{-\kappa_r(\eta)}.
\ee
For instance, for the sudden reionization model,  one easily obtains
\ba \label{kappa}
\kappa_r(\eta)=&&\frac{C_c}{3} \frac{l_H^3}{a_m^2}
     \left[(\eta-\eta_m)^{-3}-
       (\eta_E-\eta_m)^{-3}\right] \nonumber\\
&&+\frac{C_c}{2\gamma+1} l_H \left[(\eta_a-\eta_0)^{2\gamma+1}-
(\eta_a-\eta_E)^{2\gamma+1}\right], \ \ \ \ (\eta \geq \eta_{r}),
\ea
where all the  parameters have been given bellow Eq.(\ref{accel}).
For a reionization model, the most important quantity
$\kappa_r \equiv\kappa_r(\eta_b) $
is the value of the optical depth
from $\eta_0$ back up to some time $\eta_b$ before the reionization,
where $q_r(\eta_b)$ is practically vanishing.
For example, one can take $\eta_b=\eta_{split}$.
In practice, one can conveniently take $\eta_b=\eta_r$  for the sudden model,
and take $\eta_b= 0.5$ for the $\eta$-linear and $z$-linear models.
$\kappa_r$ is an integral constraint on the reionization history.
On the observational side,
based upon treatments of a sudden model,
     WMAP 1-yr  gives $\kappa_r= 0.17\pm 0.04$ \cite{Spergel03},
and  WMAP 3-yr gives  $\kappa_r= 0.09\pm 0.03$ \cite{spergel},
and  WMAP 5-yr gives $\kappa_r= 0.087\pm 0.017$ \cite{dunkley},
and  WMAP 5-yr combined with SN and BAO
yields $\kappa_r  = 0.084\pm 0.016$  \cite{komatsu,hinshaw}.
To be specific in calculation,
we will take the value $\kappa_r =0.084$ for
all three reionization models in this paper,
except when it is mentioned otherwise.
However, note that, for extended reionization models,
one should be careful in applying the WMAP observed value of $\kappa_r$,
as it is obtained by using a sudden model.
For the $\eta$-linear  model
with $X_e(\eta)$ given in Eq.(\ref{lineargradual}),
one uses the formulae of Eqs.(\ref{q}) (\ref{kappar})  (\ref{Vr})
to compute $q_r(\eta)$, $\kappa_r(\eta)$,  $V_r(\eta)$.
For the $z$-linear model with $X_e(\eta)$ in Eq.(\ref{Xeetalinear}),
one does similar computations.
The resulting $q_r(\eta)$, $\kappa_r(\eta)$,
and $V_r(\eta)$ for these three models are plotted
in Fig.\ref{fig3}.

The value of optical depth $\kappa_r$ determines
the area $\int^{\eta_0}_{\eta_{split}}V_r(\eta) d\eta$
introduced in Eq.(\ref{unitary}).
For a fixed $\kappa_r=0.084$,
the integration of  Eq.(\ref{Vr})
yields  $\int^{\eta_0}_{\eta_{split}}V_r(\eta) d\eta=0.0795$
in the sudden model,
$\int^{\eta_0}_{\eta_{split}}V_r(\eta) d\eta=0.07953$
in the $\eta$-linear model,
and $\int^{\eta_0}_{\eta_{split}}V_r(\eta) d\eta=0.07973$
in the $z$-linear model, respectively.
So two gradual models have slightly larger area than the sudden model.
Besides, our computations  also show that a larger $\kappa_r$
 yields a larger  $\int^{\eta_0}_{\eta_{split}}V_r(\eta) d\eta$
 and a smaller $\int^{\eta_{split}}_0V_d(\eta) d\eta$ due to Eq.(\ref{unitary}),
meaning that a CMB photon reaching us was more likely last scattered
at reionization.
As we shall see explicitly,  for CMB spectra,
this will enhance the reionization bumps on large scales
and reduce the primary peaks due decoupling.

\begin{figure}
\caption{\label{fig4}
The $\eta$-linear reionization model with $\kappa_r=0.084$.
The solid lines are the calculated results.
The dashed lines are the fitting by
two half Gaussian functions in Eq.(\ref{linearvisibility}).
}
\end{figure}

\begin{figure}
\caption{  \label{fig5}
The $z$-linear reionization model and its fitting.
}
\end{figure}
To facilitate analytical calculations of CMB polarization,
similar to the treatments of $V_d(\eta)$ for the decoupling,
$V_r(\eta)$ can be also approximated by some fitting formula.
For  the $\eta$-linear model,
it is fitted by
the following two pieces of half Gaussian functions
\ba \label{linearvisibility}
V_r(\eta) =
          \left\{
\begin{array}{ll}
V(\eta_{r})\exp\left(-\frac{(\eta-\eta_{r})^2}
  {2(\Delta \eta_{r1})^2}\right), & \ \ \ \ \ \ (\eta<\eta_{r}), \\
V(\eta_{r})\exp\left(-\frac{(\eta-\eta_{r})^2}
  {2(\Delta \eta_{r2})^2}\right), &  \ \ \ \ \ \
         ( \eta >\eta_{r}),
\end{array}
\right.
\ea
where
$\Delta \eta_{r1} =0.147$, $\Delta \eta_{r2} =0.425$,
 $\Delta \eta_r=(\Delta \eta_{r1}+\Delta \eta_{r2})/2=0.286$,
and $\eta_{r}=0.935$ ($z_{r}=10.5$).
It is plotted in Panel (c) of Fig.\ref{fig4}
under the requirement that it gives the same area
$\int^{\eta_0}_{\eta_{split}}V_r(\eta) d\eta$
as the calculated one.
For the  $z$-linear model,
the fitting formula is similar to Eq.(\ref{linearvisibility})
but with the parameters
$\Delta \eta_{r1} =0.100$,  $\Delta \eta_{r2} =0.366$,
$\Delta \eta_r=(\Delta \eta_{r1}+\Delta \eta_{r2})/2=0.233$,
and $\eta_r=0.855$ $(z_r=13)$.
It is plotted in Panel (c) of Fig. \ref{fig5}.
Here for the two extended models,
the value of $\eta_r$ has been taken
to correspond to the maximum of $V_r(\eta)$.
For the sudden model,
it can be fitted by a half piece of Gaussian function
\ba \label{vreion}
V_r(\eta) =
          \left\{
\begin{array}{ll}
0, & \ \ \ \ \ \ \ \ \ ( {\rm for } \,\,\eta<\eta_{r}), \\
 V(\eta_{r})\exp\left(-\frac{(\eta-\eta_{r})^2}
  {2(\Delta \eta_{dr})^2}\right), &  \ \ \ \ \ \ \ \ \
         (  {\rm for }  \,\, \eta >\eta_{r}),
\end{array}
\right.
\ea
with the width  $\Delta \eta_{dr}=0.247$,
plotted in Panel (c) of Fig.\ref{fig6}.
The half-gaussian
fitting of $V_r(\eta)$ for the sudden model
is not as accurate  as those for the two extended models.
It should be expected that in the sudden model
the analytical CMB spectra $C_l^{XX}$
based on its fitting formula (\ref{vreion})
is not as good as those in the two extended models.

We mention that, given a fixed $\kappa_r$,
the respective height $V(\eta_r)$
in Eqs.(\ref{sudden}), (\ref{vreion}), and (\ref{linearvisibility})
 are also determined automatically.
From these  fitting $V_r(\eta)$,
one can convert it to obtain the corresponding optical functions
\be \label{rkappa}
e^{-\kappa_r(\eta)}=  1-\int^{\eta_0}_{\eta} V_r(\eta) d \eta ,
\ee
\be \label{kappafit}
\kappa_r(\eta)=
     -\ln \left(1-\int^{\eta_0}_{\eta} V_r(\eta) d \eta \right),
\ee
\be \label{qfit}
q_r(\eta)=\frac{V_r(\eta)}{\left(1-\int^{\eta_0}_{\eta}
         V_r(\eta) d \eta \right)}\, \,.
\ee
It should be mentioned that
the approximate fitting of $V_r(\eta)$ by Eq.(\ref{linearvisibility})
underestimates the value of $V_r$  in the range $\eta>\eta_r$
by $\sim 9.1\% $.
For the $z-$linear model, the fitting by  half Gaussian functions
underestimates the value of $V_r$  in the range $\eta>\eta_r$
by $\sim 8.6\%$.
However, this kind of error of the fitting
can partially compensated in treating the damping factors
occurring in the time integration of the polarization mode,
as will be given in the following.
The gaussian fitting of Eq.(\ref{vreion}) for the sudden model
is included only for illustration purpose,
as its error is larger than the two extended  models.

\begin{figure}
\caption{\label{fig6}
The sudden reionization model and its fitting.
Panel (c) shows that the fitting $V_r(\eta)$ by Eq.(\ref{vreion})
has large errors to the calculated one.
The evolution history of $V(\eta)$,
including both reionization and decoupling,
is sketched in Panel (d).
}
\end{figure}

\begin{center}
{\em\Large 4. Spectra of CMB Anisotropies and Polarization}
\end{center}

By applying the same kind of approximate integration technique
as in Refs.\cite{Zhao06,Xia},
up to the second order of a small $1/q^2$ in the tight coupling limit,
the function $G(\eta)$ in Eq.(\ref{G}) can be written as
\be \label{G0}
G(\eta)
=  - \frac{1}{10}   \int_0^{\eta} \dot h (\eta')\,
e^{-\frac{3}{10}\kappa (\eta')-\frac{7}{10}\kappa (\eta)}d \eta',
\ee
and
the integration of polarization mode in Eq.(\ref{betae}) is written as
\be
\beta_l(\eta_0) =   -\frac{1}{10}i^l
\int_0^{\eta_0} d\eta V(\eta)\dot{h}(\eta)j_l(k(\eta-\eta_0))
\int_0^{\eta} d\eta' e^{-\frac{3}{10}\kappa(\eta')-\frac{7}{10}\kappa(\eta)}.
\ee
Since the visibility function $V(\eta)$ for the whole history
consists of
two effectively non-overlapping functions, $V_d(\eta)$ and $V_r(\eta)$,
the $\eta$-time integration $\int_0^{\eta_0} d\eta$
in the above
is naturally split into a sum of two integrations:
\ba \label{betaspl}
\beta_l(\eta_0) =  -\frac{1}{10}i^l
\int_0^{\eta_{split}} d\eta V_d(\eta)\dot{h}(\eta)j_l(k(\eta-\eta_0))
\int_0^{\eta} d\eta' e^{-\frac{3}{10}\kappa(\eta')-\frac{7}{10}\kappa(\eta)}\nonumber \\
  -\frac{1}{10}i^l
\int_{\eta_{split}}^{\eta_0} d\eta V_r(\eta)\dot{h}(\eta)j_l(k(\eta-\eta_0))
\int_0^{\eta} d\eta' e^{-\frac{3}{10}\kappa(\eta')-\frac{7}{10}\kappa(\eta)}.
\ea
One defines  the integration  variable
$x\equiv \kappa(\eta')/ \kappa(\eta)$ to replace the variable $\eta'$ in the above.
Since  $V_d(\eta)$ is peaked around $\eta_d$ with a width $\Delta\eta_d$,
and, similarly, $V_r(\eta)$ is peaked around $\eta_r$ with a width $\Delta\eta_r$,
one can take $d\eta' \simeq -\Delta\eta_d \frac{dx}{x}$ and
$d\eta' \simeq -\Delta\eta_r \frac{dx}{x}$ as approximation, respectively.
\ba \label{betax}
\beta_l(\eta_0) =  -\frac{1}{10}i^l
\Delta\eta_d\int_0^{\eta_{split}} d\eta V_d(\eta)\dot{h}(\eta)j_l(k(\eta-\eta_0))
\int_1^{\infty} \frac{dx}{x} e^{-\frac{3}{10}\kappa(\eta)x-\frac{7}{10}\kappa(\eta)}\nonumber \\
  -\frac{1}{10}i^l
\Delta\eta_r\int_{\eta_{split}}^{\eta_0} d\eta V_r(\eta)\dot{h}(\eta)j_l(k(\eta-\eta_0))
\int_1^{\infty} \frac{dx}{x} e^{-\frac{3}{10}\kappa(\eta)x-\frac{7}{10}\kappa(\eta)}.
\ea
For each term in the above,
the $\eta$-time integration can be dealt with,
using the same kind of treatment as in Ref.\cite{Zhao06,Xia}.
For the decoupling one has
\be \label{intd}
\int_0^{\eta_{split}} d\eta V_d(\eta)\dot{h}(\eta)j_l(k(\eta-\eta_0))
\simeq D_d(k)\dot{h}(\eta_d)j_l(k(\eta_d-\eta_0))
        \int_0^{\eta_{split}} d\eta V_d(\eta),
\ee
where the damping factor for the decoupling is given by
the following fitting formula
\be \label{D}
D_d(k)=\frac{1.4}{2}
[e^{-c(k\Delta\eta_{d1})^b}+e^{-c(k\Delta\eta_{d2})^b}],
\ee
which can be simplified by
\be \label{d}
D_d(k)=1.4 e^{-c(k\Delta\eta_{d})^b},
\ee
with $c$ and $b$ being two fitting parameters.
For CMB spectra without reionization,
it has been shown in Ref.\cite{Xia} that
both damping factors in Eqs.(\ref{D}) and (\ref{d})
$c\simeq 0.6$ and $b\simeq 0.85$
give a good match with the numerical result by CAMB  \cite{Lewis}
over an extended range $l\leq 600$,
covering the first three primary peaks,
and the error is only $\sim 3\%$.

Similarly, the $\eta$-time integration for the  reionization is
\be \label{intr}
\int_{\eta_{split}}^{\eta_0} d\eta V_r(\eta)\dot{h}(\eta)j_l(k(\eta-\eta_0))
\simeq D_r(k)\dot{h}(\eta_r)j_l(k(\eta_r-\eta_0))
       \int_{\eta_{split}}^{\eta_0}d\eta V_d(\eta),
\ee
where the damping factor for the extended models is taken to be
\be \label{Dr}
D_r(k)=\frac{1.4}{2}
[e^{-c(k\Delta\eta_{r1})^b} + e^{-c(k\Delta\eta_{r2})^b}],
\ee
or for the sudden reionization
\be \label{dr}
D_r(k)=\frac{1.4}{2} e^{-c(k\Delta \eta_{r})^b}.
\ee
Here the parameter $c$ and $b$  in Eqs.(\ref{Dr}) and (\ref{dr})
for reionization
could take values different from those for decoupling.
For simplicity,
we let them take the values that are the same as in $D_d(k)$.
Guided by the error estimation for the decoupling case,
we can only estimate the errors due to $D_r(k)$
in Eq.(\ref{Dr}) for the two extended  models
upon the reionization bumps of polarization spectra to be $\le 10\%$,
the same order of magnitude
as those of the fitting $V_r(\eta)$ in Eq.(\ref{linearvisibility}).

Substituting Eqs.(\ref{intd}) and (\ref{intr}) into Eq.(\ref{betax}),
and performing the integrations $\int d\eta$ first,
\be
\int_0^{\eta_{split}}   d\eta V_d(\eta)
       e^{-\frac{3}{10}\kappa(\eta)x-\frac{7}{10}\kappa(\eta)}
=  \int_{\kappa_r}^\infty d\kappa
       e^{-\frac{3}{10}\kappa x-\frac{17}{10}\kappa}
=\frac{1}{\frac{17}{10}+\frac{3}{10}x}e^{-(\frac{17}{10}+\frac{3}{10}x)\kappa_r }
\ee
\be
\int_{\eta_{split}}^{\eta_0}   d\eta V_r(\eta)
       e^{-\frac{3}{10}\kappa(\eta)x-\frac{7}{10}\kappa(\eta)}
=  \int_0^{\kappa_r} d\kappa
       e^{-\frac{3}{10}\kappa x-\frac{17}{10}\kappa}
=\frac{1}{\frac{17}{10}+\frac{3}{10}x}
   \left[ 1-e^{-(\frac{17}{10}+\frac{3}{10}x)\kappa_r } \right]
\ee
one finally obtains the expression of the polarization mode
as a sum of two parts
\ba \label{betar2}
\beta_l(\eta_0) = -\frac{1}{10}
i^l &&\left[   A_1(\kappa_r) D_d(k) \Delta \eta_d
   \dot{h}(\eta_d)  j_l(k(\eta_d-\eta_0))  \right. \nonumber\\
&&\left.  +A_2(\kappa_r) D_r(k) \Delta \eta_{r}
    \dot{h}(\eta_{r})  j_l(k(\eta_r-\eta_{0})) \right]
\ea
where the $\kappa_r$-dependence coefficients
\be \label{A1}
A_1(\kappa_r) = \int_1^\infty \frac{dx}{x (\frac{17}{10}+\frac{3}{10}x) }
                e^{-(\frac{17}{10}+\frac{3}{10}x)\kappa_r},
\ee
\be \label{A2}
A_2(\kappa_r) = \int_1^\infty \frac{dx}{x (\frac{17}{10}+\frac{3}{10}x) }
                \left[1-e^{-(\frac{17}{10}+\frac{3}{10}x)\kappa_r}\right],
\ee
both being independent of the wavenumber $k$,
and the sum is
$A_1(\kappa_r) +A_2(\kappa_r)=\frac{10}{17}\ln \frac{20}{3}\simeq 1.116$,
independent of $\kappa_r$.
If one sets $A_2=0$ and   $A_1=\frac{10}{17}\ln \frac{20}{3}$,
Eq.(\ref{betar2}) reduces to exactly that of the non-reionization case
\cite{Zhao06,Xia}.
Actually, after the sum is normalized to unity,
the two coefficients have the  physical meaning:
\be\label{a1}
a_1(\kappa_r)\equiv
\frac{A_1(\kappa_r)}{\frac{10}{17}\ln \frac{20}{3}}
\ee
is the probability that a polarized photon we perceive was
last scattered during the decoupling epoch,
and
\be \label{a2}
a_2(\kappa_r)\equiv
\frac{A_2(\kappa_r)}{\frac{10}{17}\ln \frac{20}{3}}
\ee
is the probability that a polarized photon we perceive
was last scattered during the time interval
from the beginning of reionization up to the present time $\eta_0$.
It is found that
$a_1(\kappa_r)$ is a decreasing function of  $\kappa_r$
and $a_2(\kappa_r)$ is an increasing one,
as shown in Fig. \ref{fig7}.
Therefore,
if more CMB photons are scattered
by the free electrons during the reionization,
the optical depth $\kappa_r$ acquires a larger value,
giving rise to a higher coefficient $A_2(\kappa_r)$
and, at the same time, a lower coefficient $A_1(\kappa_r)$.
The $A_1(\kappa_r)$ part in $\beta_l$ from the decoupling
will give rise to the primary peaks of $C_l^{EE}$ and  $C_l^{BB}$ ,
and will be prominent on small angular scales with $l\ge 100$.
The $A_2(\kappa_r)$ part from the reionization will be dominant
on large angular scales and will yield the reionization bumps
of $C_l^{EE}$ and $C_l^{BB}$ around $l<10$.

The analytical expression (\ref{betar2}) has the merit
that effects of relevant physical elements upon the polarization
have been explicitly isolated  and displayed.
The $\kappa_r$-dependence of $\beta_l$ is attributed
to the coefficients $A_1(\kappa_r)$ and $A_2(\kappa_r)$,
which determine the relative heights of the primary peaks
and the reionization bump.
Other effects of reionization is encoded
in the factor $D_r(k)\Delta\eta_r$.
The  effects of decoupling are absorbed in $D_d(k)\Delta\eta_d$.
The effect of RGWs upon the polarization
are given by the time derivatives $\dot{h}(\eta_d) $ at $\eta_d$
and $ \dot{h}(\eta_{r})$ at $\eta_r$,
which not only contain the cosmological information,
such as inflation and NFS, etc.,
more importantly,
but also determine the overall profiles of $C_l^{EE}$ and $C_l^{BB}$,
such as the locations of peaks and troughs, and of bumps.
The factors $ j_l(k(\eta_d-\eta_0))$ and $j_l(k(\eta_r-\eta_{0}))$
just play the role of conversion from the wavenumber $k$-space
into the multipole $l$-space.
\begin{figure}
\caption{\label{fig7}
The normalized coefficients $a_1(\kappa_r)$ and $a_2(\kappa_r)$
of the polarization $\beta_l$.
A larger $\kappa_r$ yields lower $a_1(\kappa_r)$
and higher $a_2(\kappa_r)$,
i.e., will yield
 lower primary peaks and higher reionization bump
in $C_l^{EE}$ and $C_l^{BB}$.
Also plotted are the coefficients
$e^{-\kappa_r}$ and  $(1-e^{-\kappa_r})$
of the temperature anisotropies $\alpha_l$.
Notice that $a_1(\kappa_r)$ and $a_2(\kappa_r)$ vary with $\kappa_r$
more drastically than $e^{-\kappa_r}$ and  $(1-e^{-\kappa_r})$,
respectively.
}
\end{figure}

To calculate the temperature anisotropies,
we need to evaluate $\xi_l$ in Eq.(\ref{xigeneral}),
which contains the factor $e^{-\kappa(\eta)}$.
This also needs to be dealt with  properly.
As shown in  Fig.\ref{fig6},
the factor $e^{-\kappa(\eta)}$
has two steps, one at the decoupling $\eta = \eta_d$,
and another at $\eta  \simeq \eta_{r}$ caused by the reionization.
It can be approximated by the following two-step function
\ba \label{ekappa}
e^{-\kappa(\eta)} \simeq
          \left\{
\begin{array}{ll}
0 & \ \ \ \ \ \ \ \ \ (\eta<\eta_{d}); \\
e^{-\kappa_r} &  \ \ \ \ \ \ \ \ \
         (\eta_{d}<\eta<\eta_{r}); \\
1 &   \ \ \ \ \ \ \ \ \   (\eta_{r}< \eta<\eta_{0}),
\end{array}
\right.
\ea
and its reionization-relevant part $e^{-\kappa_r(\eta)}$
 is between $(\eta_d,\eta_0)$.
By Eq.(\ref{rkappa}),  $e^{-\kappa_r(\eta)}$
is the integration of $V_r(\eta)$  from $\eta$ to $\eta_0$,
determined by the area under the curve of $V_r(\eta)$,
not very sensitive to the detailed shape of $V_r(\eta)$.
Therefore,
the approximate formula (\ref{ekappa}) will be used
for the three models of reionization,
with their respective values of $\eta_r$.
Note that Eq.(\ref{ekappa}) tends to  overestimate
the contribution of the reionization to the integration,
since $e^{-\kappa(\eta)}$ shown in Fig.\ref{fig6}
increases gradually from $e^{-\kappa_r}$ at $\eta_r$
up to $1$ for $\eta \gg \eta_{r}$,
instead of instantaneously jumping up to $1$ at  $\eta_{r}$.
To compensate this overestimation,
in actually calculating $\xi_l(\eta_0)$
 in the linear model,
we may use the value of $\eta_r $ slightly greater than $0.935$.
But this adjustment of the time $\eta_r$ does not apply
to $\beta_l(\eta_0)$ in Eq.(\ref{betar2}).
Substituting  Eq.(\ref{ekappa}) into  Eq.(\ref{xigeneral}),
the integration for $\xi_l$ is split into two terms
\be \label{xildr}
\xi_l(\eta_0)\simeq
i^l \int^{\eta_{r}}_{\eta_d}  e^{-\kappa_r}
\dot{h}(\eta)j_l(k(\eta_0-\eta))d \eta
+i^l \int^{\eta_0}_{\eta_{r}}
\dot{h}(\eta)j_l(k(\eta_0-\eta)) d \eta.
\ee
Following the similar treatments in \cite{baskaran,Xia},
each term is integrated by parts,
yielding  the following approximate expression
\ba \label{xirl}
\xi_l(\eta_0)= -i^l \left[
 e^{-\kappa_r}h(\eta_d)j_l(k(\eta_0-\eta_d))
+(1-e^{-\kappa_r})h(\eta_r)j_l(k(\eta_0-\eta_{r}))
\right],
\ea
where the first term is generated by $h(\eta_d)$ at the recombination
and the second term is due to $h(\eta_r)$ at the reionization.
Eq.(\ref{alpha}) then yields
the mode of CMB temperature anisotropies
$\alpha_l(\eta_0)=\xi_l(\eta_0)-\beta_l(\eta_0)$.
In fact,  $\alpha_l(\eta_0)$ is essentially contributed by  $\xi_l(\eta_0)$
since the amplitude of $\xi_l(\eta_0)$ is about two orders
higher than that of $\beta_l(\eta_0)$.
Writing down explicitly,
one has the approximate, analytic expression of
the mode of CMB temperature anisotropies, including the reionization,
\ba \label{alpharl}
\alpha_l(\eta_0)&=&-i^l j_l(k(\eta_{0}-\eta_d))\left[e^{-\kappa_r}h(\eta_d)
-\frac{1}{10}A_1(\kappa_r) D_d(k) \Delta \eta_d \dot{h}(\eta_d)\right] \nonumber\\
&& -i^l j_l(k(\eta_0-\eta_{r}))
\left[(1-e^{-\kappa_r})h(\eta_{r})
-\frac{1}{10} A_2(\kappa_r) D_r(k) \Delta \eta_{dr} \dot{h}(\eta_{r})\right].
\ea
In this expression,
the first term containing $h(\eta_d)$  and $\dot h(\eta_d)$
is brought by  the decoupling and responsible for the primary peaks,
whereas the last term containing $h(\eta_r)$  and $\dot h(\eta_r)$
is brought in by reionization
and prominent on large angle scales with $l<10$.
When one sets $A_1=1$, $A_2=0$, and $e^{-\kappa_r}=1$,
Eq.(\ref{alpharl}) reduces to the results for
the case without reionization \cite{ Xia}.
The $\kappa_r$-dependence of $\alpha_l$ is mainly attributed to
the factors $e^{-\kappa_r}$ and $(1-e^{-\kappa_r})$,
while the portion containing $A_1(\kappa_r)$ and $A_2(\kappa_r)$
is the subdominant $\beta_l$.
By Eq.(\ref{rkappa}) and the definition of $\kappa_r$, on has
\be \label{expkr}
e^{-\kappa_r}=  1-\int^{\eta_0}_{\eta_b} V_r(\eta) d \eta ,
\ee
which has a physical interpretation:
the probability of a CMB photon being last scattered
during the earlier epoch before the reionization.
Since $e^{-\kappa_r} <1$ for $\kappa_r>0$,
it will cause a slight decrease in the amplitude of
the temperature anisotropies, as demonstrated in Eq.(\ref{alpharl}).
Correspondingly,
 the factor $(1-e^{-\kappa_r})$ in front of $ h(\eta_r)$ is
 \be \label{1-expkr}
1-e^{-\kappa_r}=  \int^{\eta_0}_{\eta_b} V_r(\eta) d \eta ,
 \ee
recognized as
the probability of a CMB photon being last scattered
during the time interval
from the reionization up to the present time $\eta_0$.
These foregoing probabilistic interpretations have  the parallels
in the case of CMB anisotropies generated by scalar perturbations,
where reionization also brings about a similar
exponential factors $e^{-\kappa_r}$ in the temperature anisotropies,
and a physical illustration on its appearance
is given in Ref.\cite{Griffiths2}.
It should be mentioned
that the probabilities in Eqs.(\ref{expkr}) and (\ref{1-expkr})
are respectively different from
the normalized $a_1(\kappa_r)$ and $a_2(\kappa_r)$,
the latter are for the polarized photons.
Moreover, as shonw in Fig. \ref{fig7},
$a_1(\kappa_r)$ decreases with $\kappa_r$
much faster than $e^{-\kappa_r}$  does,
and $a_2(\kappa_r)$ increases much faster than $(1-e^{-\kappa_r})$.
In this sense,
the polarization $\beta_l(\eta_0)$ is more sensitive to
$\kappa_r$ than the temperature anisotropies $\alpha_l(\eta_0)$.
Therefore, one may say that
the polarization spectra $C^{EE}_l$ are $C^{BB}_l$
are more sensitive probes into the reionization than
the temperature anisotropies spectrum $C^{TT}_l$.

With $\alpha_l$ and $\beta_l$ being ready,
one can compute  straightforwardly  the CMB spectra
caused by RGWs.
The detailed derivations have been demonstrated
in Refs. \cite{Kamionkowski,Zhao06,Xia}.
In particular, some minor misprints of the coefficients in Ref.\cite{Kamionkowski}
have been pointed out and corrected in Refs. \cite{Zhao06,Xia}.
The temperature anisotropies
\ba \label{ctt}
C^{TT}_l=\frac{1}{8\pi}\frac{(l+2)!}{(l-2)!}
\int k^2dk \left|
\frac{\alpha_{l-2}(\eta_0) }{(2l-1)(2l+1)}-
\frac{2\alpha_{l}(\eta_0)}{(2l-1)(2l+3)}
+\frac{\alpha_{l+2}(\eta_0)}{(2l+1)(2l+3)}\right|^2,
\ea
the electric type of polarization
\ba \label{cee}
C^{EE}_l= \frac{1}{16\pi}
   \int k^2dk \left|\frac{(l+1)(l+2)\beta_{l-2}(\eta_0)}{(2l-1)(2l+1)}+
\frac{6(l-1)(l+2)\beta_{l}(\eta_0)}{(2l-1)(2l+3)}+\frac{l(l-1)
   \beta_{l+2}(\eta_0)}{(2l+1)(2l+3)} \right|^2 ,
\ea
the magnetic type of polarization
\ba \label{cbb}
C^{BB}_l = \frac{1}{16\pi}\int k^2dk
     \left|\frac{2(l+2)\beta_{l-1}(\eta_0)}{(2l+1)}
     +\frac{2(l-1)\beta_{l+1}(\eta_0)}{(2l+1)} \right|^2 ,
\ea
and the temperature-polarization cross
\ba \label{cte}
C^{TE}_l&=&\sqrt{\frac{1}{8\pi}\frac{(l+2)!}{(l-2)!}}
\sqrt{\frac{1}{16\pi}} \int k^2dk
\left[  \frac{\alpha_{l-2}(\eta_0) }{(2l-1)(2l+1)}
        - \frac{2 \alpha_{l}(\eta_0)}{(2l-1)(2l+3)}
        +\frac{\alpha_{l+2}(\eta_0)}{(2l+1)(2l+3)} \right]
\nonumber\\ &&\times
\left[\frac{(l+1)(l+2)\beta_{l-2}(\eta_0)}{(2l-1)(2l+1)}+
\frac{6(l-1)(l+2)\beta_{l}(\eta_0)}{(2l-1)(2l+3)}+\frac{l(l-1)
   \beta_{l+2}(\eta_0)}{(2l+1)(2l+3)} \right] .
\ea
Substituting $\alpha_l(\eta_0)$
and $\beta_l(\eta_0)$ into Eqs. (\ref{ctt}), (\ref{cee}),
(\ref{cbb}) and (\ref{cte}) yields
the analytical expressions of the spectra of CMB
with the modifications of reionization:
\ba \label{cttr}
C^{TT}_l&=&\frac{1}{8\pi}\frac{(l+2)!}{(l-2)!}\int k^2dk
\left\{P_{Tl}(k(\eta_{0}-\eta_d))
\left[e^{-\kappa_r}h(\eta_d)
-\frac{1}{10}
A_1 D_d(k) \Delta \eta_d \dot{h}(\eta_d)\right]
\right.\nonumber\\
&&\left.+P_{Tl}(k(\eta_0-\eta_{r}))
\left[(1-e^{-\kappa_r})h(\eta_{r}) -\frac{1}{10}
A_2 D_r(k) \Delta \eta_{r} \dot{h}(\eta_{r})\right] \right\}^2,
\ea
\ba \label{ceer}
C^{EE}_l
&=&\frac{1}{16\pi}\left(\frac{1}{10}\right)^2
\int k^2dk \left[P_{El}(k(\eta_0-\eta_d))A_1 D_d(k)
\Delta \eta_d \dot{h}(\eta_d)\right. \nonumber\\
&&\left.+P_{El}(k(\eta_0-\eta_{r}))A_2 D_r(k)
\Delta \eta_{r} \dot{h}(\eta_{r})\right]^2,
\ea
\ba \label{cbbr}
C^{BB}_l
&=&\frac{1}{16\pi}\left(\frac{1}{10} \right)^2
\int k^2dk \left[P_{Bl}(k(\eta_0-\eta_d))A_1 D_d(k)
\Delta \eta_d \dot{h}(\eta_d)\right. \nonumber\\
&&\left.+P_{Bl}(k(\eta_0-\eta_{r}))A_2 D_r(k)
\Delta \eta_{r} \dot{h}(\eta_{r})\right]^2,
\ea
\ba \label{cter}
C^{TE}_l
&=&-\frac{1}{8\sqrt{2}\pi\times 10 }
\sqrt{\frac{(l+2)!}{(l-2)!}}
\int k^2dk  \nonumber\\
&&\frac{1}{2}\left\{ \left[ P_{Tl}(k(\eta_{0}-\eta_d))
\left[e^{-\kappa_r}h(\eta_d)
-\frac{1}{10} A_1 D_d(k) \Delta \eta_d \dot{h}(\eta_d) \right] \right.\right.
\nonumber\\
&&\left.\left. +P_{Tl}(k(\eta_0-\eta_{r}))
\left[(1-e^{-\kappa_r})h(\eta_{r}) -\frac{1}{10}
A_2 D_r(k) \Delta \eta_{dr} \dot{h}(\eta_{r})\right] \right ]\right. \nonumber\\
&&\left.\left[P_{El}(k(\eta_0-\eta_d))A_1 D_d(k)
\Delta \eta_d \dot{h}(\eta_d)
+P_{El}(k(\eta_0-\eta_{r}))A_2 D_r(k)
\Delta \eta_{r} \dot{h}(\eta_{r})\right]^* \right. \nonumber\\
&&  \left.+ {\rm Complex \,\, Conjugate}. \right\}
\ea
In the above integrations ,
the projection factors are defined as:
\ba \label{pt}
P_{Tl}(x)=\frac{j_{l-2}(x)}{(2l-1)(2l+1)}+
         \frac{2j_{l}(x)}{(2l-1)(2l+3)}+\frac{j_{l+2}(x)}{(2l+1)(2l+3)}
         =\frac{j_l(x)}{x^2},
\ea
\ba  \label{pe}
P_{El}(x)
&&=\frac{(l+1)(l+2)}{(2l-1)(2l+1)}j_{l-2}(x)
    -\frac{6(l-1)(l+2)}{(2l-1)(2l+3)}j_{l}(x)+\frac{l(l-1)}
{(2l+1)(2l+3)}j_{l+2}(x) \nonumber \\
&&=-[2-\frac{l(l-1)}{x^2}]j_l(x)+\frac{2}{x}j_{l-1}(x)
\ea
\be \label{pb}
P_{Bl}(x)
=\frac{2(l+2)}{(2l+1)}j_{l-1}(x)
       -\frac{2(l-1)}{(2l+1)}j_{l+1}(x) \nonumber \\
=2j_{l-1}(x)-2\frac{l-1}{x}j_l(x).
\ee
We apply these formulae to
the three reionization models, respectively,
and  plot the spectra $C_l^{XX}$.
The reionized spectra  $C_l^{XX}$ are plotted in Fig. \ref{fig8}
for the three models of reionization,
in which we also plot the numerical spectra from the CAMB Online Tool
for a comparison \cite{Lewis}.
Both the analytic and numerical computation
use the same set of parameters  $\kappa_r=0.084$ and $r=0.37$.
On large scales $l\le 600$ our analytical  $C_l^{EE}$ and  $C_l^{BB}$
agree with the numerical ones.
For the two extended  models,
the error is $\sim 3\%$ for the primary peaks,
and the error is estimated to be $\le 15\%$ for the reionization bumps
as superposed from that of decoupling $\sim 3\%$
and that of reionization $\sim 10\%$.
Notice also that the analytical $C_l^{EE}$ and $C_l^{BB}$
in the sudden model have reionization bumps too low.
This has been expected,
since the half-gaussian fitting formula (\ref{vreion}) is poor.
The analytical $C_l^{TT}$ and  $C_l^{TE}$
are close to the numerical ones on smaller scales $l> 20$,
but have obvious departure from the numerical ones
on very large scales $l< 10$.
This implies that the approximation of
temperature anisotropies $\xi_l$ in  Eq.(\ref{xirl}) is
poor for small multipoles $l < 10$.
In the following we focus only on
the two extended models
and examine the impact of reionization
through the analytical spectra $C_l^{XX}$.
\begin{figure}
\caption{
\label{fig8}
The spectra $C_l^{XX}$  in three reionization models.
The parameters $c=0.65$ and $b=0.85$ in $D_r(k)$ are taken.
The optical depth  $\kappa_r=0.084$
and the ratio $r=0.37$ are taken.
The numerical result is obtained with the same set of parameters,
using  CAMB  \cite{Lewis}.
}
\end{figure}

\begin{center}
{\em\Large 5. Effects of Reionization }
\end{center}

1.
The most prominent modification due to the reionization
is that it enhances the low-$l$ parts of the spectra,
forming a reionization bump at $l\sim 5$ for $C^{EE}_l$ and $C^{BB}_l$,
respectively.
The position of this bump
is a reflection of the horizon scale at reionization,
whose corresponding angular scale, $l\sim 5$,
is much larger than $l \sim 100$
of the primary peaks at the photon decoupling.
As pointed out earlier, the  profiles of  $C^{XX}_l$
are determined by the profiles of RGWs at the decoupling
and at the reionization as well.
In particular,
the reionization bumps are generated by
$\dot{h}(\eta_r)$,
and the primary peaks and troughs
are due to $h(\eta_d)$ and $\dot{h}(\eta_d)$.
This correspondence is clearly demonstrated  by Fig.\ref{fig9},
in which the left panel plots $C^{EE}$ and $C^{BB}$,
as well as  $\dot{h}(\eta_d)$ and $\dot{h}(\eta_r)$ in one graph,
and the right panel plots $C^{TT}$,
as well as $h(\eta_d)$ and $h(\eta_r)$ in one graph.
This correspondence can be further explained
by the following analysis.
The respective projection factors,  $P_{Tl}$, $P_{El}$, and $P_{Bl}$
as the integrands of $C^{XX}_l$
are made up of the spherical  Bessel's  functions,
$j_l(x)$, which is sharply peaked around $x \simeq l$.
Subsequently the projection factors as functions of $k$
are sharply peaked around
\be \label{k-l}
k(\eta_0-\eta_d)  \simeq k \eta_0\simeq l,
\ee
\be \label{k-lr}
k(\eta_0-\eta_r) \simeq l,
\ee
respectively.
Consequently,  the spectra as  integrations  over $k$
will receive  main contributions from
the integration range $k\sim l/\eta_0$ to the primary peaks
and from $k\sim l/(\eta_0-\eta_r)$ to the bump, respectively  \cite{Zhao06}:
\be \label{cee-appr}
C_l^{EE}, \,  C_l^{BB}
\propto
  A_1^2D_d^2(k)\left|\dot{h}(\eta_d)\right|^2_{k \sim l/\eta_0}
+ A_2^2D_r^2(k)\left|\dot{h}(\eta_r)\right|^2_{k \sim l/(\eta_0-\eta_r)} ,
\ee
\be \label{ctt-appr}
C^{TT}_l \propto
e^{-2\kappa_r} \left|h(\eta_d)\right|^2_{k \sim l/\eta_0}
+(1-e^{-\kappa_r} )^2  \left|h(\eta_r)\right|^2 _{k \sim l/(\eta_0-\eta_r)},
\ee
\be \label{cte-appr}
C^{TE}_l \propto
 A_1 e^{-\kappa_r} D_d(k) h(\eta_d) \dot{h}(\eta_d)_{k \sim l/\eta_0}
+ A_2(1-e^{-\kappa_r}) D_r(k)
   h(\eta_r)\dot{h}(\eta_r)_{k \sim l/(\eta_0-\eta_r)}.
\ee
According to Eq.(\ref{cee-appr}),
the locations of the primary peaks of $C_l^{EE}$ and $C_l^{BB}$
are mainly determined by the  $|\dot{h}(\eta_d)|^2$-term,
and those of the reionization bumps are determined
by the $|\dot{h}(\eta_r)|^2$-term.
However, the spectrum $C_l^{TT}$
does not have a prominent bump around $l\sim 5$.
This is because both $|h(\eta_d)|^2$ and $|h(\eta_r)|^2$
have a similar slope around there,
and their superposition only enhances the spectral amplitude,
not forming a bump.
These are illustrated in Fig. \ref{fig9}.
\begin{figure}
\caption{
\label{fig9}
The correspondence of the profiles of $C_l^{XX}$
with that of RGWs at the decoupling and at the reionization.
$\dot{h}(\eta_r)$  is responsible
for the bumps of $C_l^{EE}$ and  $C_l^{BB}$ around $l\sim5 $,
while $\dot{h}(\eta_d)$
is responsible the primary peaks and troughs for $l\geq 100$.
For $C_l^{TT}$,
$h(\eta_d)$ and $h(\eta_r)$ have a similar slope at $l\le 10$,
and their superposition does not form a prominent bump of $C_l^{TT}$.
}
\end{figure}

2.
The reionization bumps in the polarization spectra
depend on the detailed reionization history.
$C_l^{EE}$ and $C_l^{BB}$ for a fixed value of the optical depth
$\kappa_r=0.084$ in the two extended models are
shown in Fig. \ref{fig10}.
The bumps in the $\eta$-linear model
are located at a slightly larger angular scale (smaller $l$)
than that in the $z$-linear model.
This is because we have assigned
a greater $\eta_r=0.935$ in the $\eta$-linear model
than that $\eta_r=0.855$ in the $z$-linear model,
so its bump is located at a slightly smaller $l\sim k(\eta_0-\eta_r)$.
Notice also that
the $\eta$-linear model  produces
higher bumps than the $z$-linear model.
This is due the fact that
the $\eta$-linear model has a greater width $\Delta\eta_r=0.286$
than that $\Delta\eta_r=0.855$ in the $z$-linear model.
Thus we conclude that
the location of bump is quite sensitive to
the the reionization time $\eta_r$,
and the height of bump is sensitive
to the width $\Delta\eta_r$ of reionization process.
This feature is helpful to probe $\eta_r$ and $\Delta\eta_r$
only if observational data on the bumps are accurate enough.
However,
when we let the two models to have the same set of
parameters $\eta_r$ and $\Delta\eta_r$,
their reionization bumps
predicted by our analytical formulation are very similar.
The lesson is that the bump is an integrating result
from the ionization fraction $X_e(\eta)$,
and, in this regards,
two different reionization histories via $X_e(\eta)$
can lead to similar bumps,
as long as they have similar  $V_r(\eta)$ \cite{Colombo1,Mukherjee}.
\begin{figure}
\caption{   \label{fig10}
$C_l^{EE}$ and  $C_l^{BB}$ in the extended reionization models.
The two models yield different reionization  bumps at $l\sim 5$
since they are assigned with
different values of $\eta_r$ and $\Delta\eta_r$.
}
\end{figure}

3. The overall profiles of CMB spectra are very
sensitive to the optical depth $\kappa_r$ of reionization.
In particular,
$\kappa_r$ is strongly degenerate with the normalization
of the amplitude of primordial fluctuations,
and this fact has been one of main difficulties
to probe the details of reionization process
\cite{MZaldarriaga,ZSS,ZS,Venkatesan1,Venkatesan2,
Griffiths,Kaplinghat,Colombo3}.
It should be emphasized that the reionization does not change
the primordial amplitude $A$ of RGWs in Eq.(\ref{initialspectrum}),
which is implicitly contained in $h(\eta)$ and $\dot{h}(\eta)$.
The impact of  $\kappa_r$ is through
the coefficients $A_1(\kappa_r)$
and $A_2(\kappa_r)$ in $\beta_l$ in Eq.((\ref{betar2})),
as well as the coefficients $e^{-\kappa_r}$
and $(1-e^{-\kappa_r})$ in $\alpha_l$ in Eq.(\ref{alpharl}).
The main features of the  $\kappa_r-A$ degeneracy are clearly revealed
by the analytical estimations
in Eqs.(\ref{cee-appr}), (\ref{ctt-appr}), and (\ref{cte-appr}).

For instance, look at Eq.(\ref{cee-appr})
for $C_l^{EE}$ and  $C_l^{BB}$.
A larger $\kappa_r  $ gives smaller $A_1$ and larger $A_2$,
leading to lower primary peaks and higher bumps of
$C_l^{EE}$ and  $C_l^{BB}$,
as illustrated in Fig. \ref{fig11}.
But, this lowering of primary peaks can be compensated
by  a choice of a higher amplitude normalization $A$,
which enhances the amplitude of $\dot h(\eta_d)$,
resulting in the unchanged term $A_1^2(\kappa_r)|\dot h(\eta_d)|^2$,
so that the primary peaks  remain the same.
This is the  $\kappa_r-A$ degeneracy.
Similar degeneracy in $C_l^{TT}$ and $C_l^{TE}$
are also understood by  Eq.(\ref{ctt-appr}) and Eq.(\ref{cte-appr}).

The $\kappa_r-A$ degeneracy can be broken.
Again, take $C_l^{EE}$ and  $C_l^{BB}$ as an example.
While a larger $\kappa_r$ and a higher $A$
can yield the unchange primary peaks,
the reionization bumps get doubly enhanced,
since the bump term  $A_2^2(\kappa_r)|\dot h(\eta_r)|^2$
in Eq.(\ref{cee-appr}) gets doubly enhanced.
This suggests a possible way  to break the degeneracy.
Eq.(\ref{cee-appr}) tells that
the relative height of the primary peaks and the bump is given by
\be
\frac{\rm  \,primary \, peak\, amplitude }{\rm  bump \, amplitude}
  \propto
\frac{A_1^2(\kappa_r)\left|\dot{h}(\eta_d)\right|^2}
{A_2^2(\kappa_r) \left|\dot{h}(\eta_r)\right|^2}.
\ee
For any given RGWs,  the ratio
$\left|\dot{h}(\eta_d)\right|/\left|\dot{h}(\eta_r)\right|$
is independent of $A$ and completely determined,
so one has
\be
\frac{\rm  \,primary \, peak\, amplitude }{\rm  bump \, amplitude}
  \propto \left(\frac{A_1(\kappa_r)}{A_2(\kappa_r)}\right)^2.
\ee
This ratio only depends on the value of $\kappa_r$
and is not sensitive to the details of a reionization model.
Therefore, using this ratio of heights,
one can infer the value of $\kappa_r$
from the  observational data of $C_l^{EE}$ and  $C_l^{BB}$,
thus breaking the degeneracy.
\begin{figure}
\caption{
\label{fig11}
The $\kappa_r-A$ degeneracy.
A larger  value  of $\kappa_r$ enhances
the bumps at $l\sim 5$ and, at the same time,
reduces the primary peaks of $C_l^{EE}$ and $C_l^{BB}$.
The plot is made for the $z$-linear model.
This degeneracy behavior also exists in
the $\eta$-linear model.
}
\end{figure}

4. The primordial fluctuation spectral index $\beta_{inf}$
introduced in Eq.(\ref{initialspectrum})
is a very important parameter for inflationary models.
Given a normalization $A$ of the RGWs amplitude,
a large $\beta_{inf}$  tilts the spectrum $h(\nu, \eta_i)$,
in such a way that RWGs is more strongly enhanced on smaller scales
 \cite{zhang06b,Miao07}.
The RGWs-generated spectra $C_l^{XX}$
are subsequently tilted in the same way \cite{Zhao06,Xia}.
Therefore, a larger $\beta_{inf}$ brings about
a similar effect on $C_l^{XX}$ as a smaller $\kappa_r$ does,
leading to certain bias in determining $\kappa_r$
\cite{Colombo3,Jungman,Eisenstein,Colombo,Mortonson2008}.
Take $C_l^{EE}$ and  $C_l^{BB}$ as an example,
for which the effect is more prominent.
Fig. \ref{fig12} shows that,
for the $z-$linear model,
 the case $(\beta_{inf}=-2.02, \kappa_r=0.106)$
 and the case $(\beta_{inf}=-2.10, \kappa_r=0.084)$
yield almost overlapping curves of the bump and
the $1^{st}$  primary peak as well.

The  $\kappa_r - \beta_{inf}$ degeneracy can also be understood
by the analytical estimation in Eq.(\ref{cee-appr}).
While a large $\beta_{inf}$ enhances $\left|\dot{h}(\eta_d)\right|^2$
on small scales,
a large $\kappa_r$ suppresses $A_1(\kappa_r)$,
resulting in an unchanged combination
$ A_1(\kappa_r)^2 \left|\dot{h}(\eta_d)\right|^2$
for the primary peaks.
But this degeneracy is clearly broken from the $2^{nd}$ primary peak on.
This is because the $\kappa_r$-induced change
in $A_1(\kappa_r)$ is scale-independent,
whereas the $\beta_{inf}$-induced change in $|h(\eta_d)|$
depends on the scale.
Therefore, one expects that
data of the smaller scale $C_l^{EE}$ and $C_l^{BB}$
will be helpful in breaking the $\kappa_r-\beta_{inf}$ degeneracy.
Note also that the principal component method
developed in Ref.\cite{Mortonson2008}
can protect the bias of $\kappa_r$ caused by $\beta_{inf}$.
\begin{figure}
\caption{
\label{fig12}
The $\kappa_r-\beta_{inf}$ degeneracy.
Although the bumps and the $1^{st}$ primary peaks
are degenerate,
the $2^{nd}$ and $3^{rd}$ primary peaks show
clear departure.
The plot is made for the $z$-linear model.}
\end{figure}

5.  Although the magnetic type of polarization $C_l^{BB}$
is thought to be  a ``smoking gun'' of detection of RGWs,
its detection is not done yet,
which may be accomplished by a future CMBpol experiment \cite{Baumann}.
The 5-year WMAP \cite{komatsu,nolta}
has given the observed cross-spectrum $C_l^{TE}$,
which is negative (anti-correlation) in a range $l\sim (50, 220)$.
Yet this observed $C_l^{TE}$ is a superposition of contributions
by both scalar perturbations and RGWs.
In order to extract the traces of RGWs out of $C_l^{TE}$,
one still needs to disentangle the contribution by RGWs from the total.
In the so-called zero-multipole method
\cite{Keating,baskaran,Polnarevmiller,MillerKeating},
one examines the impact of
the tensor/scalar ratio $r$ upon the zero multipole $l_0$
around $\sim 50$,
where $C_l^{TE}$ first crosses the value $0$ and turns negative.
However, there are other factors that can influence the value of $l_0$.
The variation of $l_0$
caused by NFS  has been estimated
to be small $\Delta l \le 4$ \cite{Xia}.
Here the reionization is another important factor that
brings about a change of $l_0$,
as is shown in Fig. \ref{fig13}
for the extended reionization models with $\kappa_r=0.084$.
Around the relevant region of $l\sim 50$,
the reionization shifts the curve of $C_l^{TE}$
to smaller angular scales by an amount of $\Delta l\sim 20$,
in comparison with the non-reionized $C_l^{TE}$.
This amount is much larger than
that caused by NFS.
Moreover, the shift $\Delta l$ increases with
the optical depth $\kappa_r$.
This significant effect has to be
incorporated into the zero multipole analysis
before one can make an extraction of RGWs
from the total $C_l^{TE}$.

In this procedure,
besides disentangling
the adiabatic (constant entropy) modes that are
dominant in the scaler perturbations,
one need consider the isocurvature modes
possibly existing in the cosmological plasma \cite{Bucher2000},
which can contribute to $C_l^{XX}$ \cite{Enqvist2000}.
In particular,
the isocurvature modes  contribute
positively (correlation) to the cross spectrum $C^{TE}_l$
in the range around $l\sim 100$,
in contrast to the adiabatic modes,
which contribute negatively (anti-correlation) there.
The observed $C^{TE}_l$ from WMAP has shown the anti-correlation,
and a very stringent constraint has been found on the isocurvature contribution
with the isocurvature/adiabatic ratio $\alpha_{-1} < 0.015$ at $95\%$CL
 \cite{komatsu}.
It is interesting to compare
the contributions from RGWs and isocurvature perturbations to $C_l^{TE}$.
The comparison is very sensitive to
the ratio  $\alpha_{-1}$ and the tensor/scalar ratio $r$.
Taking the upper limit $\alpha_{-1}=0.015 $  constrained from WMAP-5,
and using the CAMB Online Tool \cite{Lewis} results for isocurvature modes
of the plasma components of baryon, CDM, and neutrino,
one finds that when  $r=0.37$ is taken,
the amplitude of $C_l^{TE}$ generated by RGWs is about two orders
greater than that of the isocurvature modes.
So in this case the isocurvature can be neglected.
Only when a much smaller ratio $r=0.001$ is taken,
is the contribution by the isocurvature modes
comparable to that by RGWs.
This is demonstrated with $r=0.001$ and $r=0.01$ in Fig.\ref{fig14},
in which the numerical $C_l^{TE}$ contributed by the baryon isocurvature perturbation
has been produced from CAMB \cite{Lewis} with  $\alpha_{-1} = 0.015$.

\begin{figure}
\caption{
\label{fig13}
Reionization shifts  $C_l^{TE}$
to smaller angular scales by $\Delta l\sim 20$
around the region $l\sim 50$.
For illustration $\kappa_r=0.084$ and
$r=0.37$ have  been taken.
}
\end{figure}
\begin{figure}
\caption{
\label{fig14}
$C_l^{TE}$ by the baryon isocurvature mode is positive around $l\sim 100$,
whereas that by RGWs is negative there.
Only when a very small $r=0.001$ is taken,
is the amplitude of $C_l^{TE} $ by the isocurvature modes
comparable to that by RGWs.
The $C_l^{TE}$ by isocurvature mode is
the numerical result generated using
CAMB \cite{Lewis} with  $\alpha_{-1}=0.015$.
.
}
\end{figure}

6.  So far our analytic formulation  for reionization can only
distinguish two different extended models by comparing
their $\kappa_r$, $\eta_r$, and $\Delta\eta_r$.
The damping factor $D_r(k)$ in Eq.(\ref{Dr})
as a fitting formula
could be used to specify other fine details of
two reionization models.
Obviously, with a fixed $b$, a larger $c$ leads to lower bumps
of $C_l^{EE}$ and $C_l^{BB}$,
as shown in Fig. \ref{fig15} for the $z$-linear model.
On the other hand,
with a fixed $c$,
a larger  $b$ will yield a slightly higher reionization bumps,
while leaving the primary peaks  almost intact.
This property  can be inferred as the following.
For the reionization bump around $l\sim 5$,
the contribution is mainly from
$k\sim l/(\eta_0-\eta_r)$ according to Eq.(\ref{k-lr}) ,
so $D_r(k)\propto e^{-c(\Delta\eta_r l/(\eta_0-\eta_r))^b}$.
In the reionization models considered in this paper,
the combination $\Delta\eta_r l/(\eta_0-\eta_r) \sim 0.5<1$,
so a larger parameter $b$ leads to a larger $D_r$ and higher bumps.
For the primary peaks with $l\ge 100$,
$D_r(k)$ is so small that the term
$A_2^2D_r^2(k)\left|\dot{h}(\eta_r)\right|^2_{k \sim l/(\eta_0-\eta_r)}$
to $C_l^{EE}$ and  $C_l^{BB}$
is practically negligible,
leaving the primary peaks intact
under a variation of $b$ in $D_r(k)$.
\begin{figure}
\caption{
\label{fig15}
The damping factor $D_r(k)$ in Eq.(\ref{Dr})
depends on the parameters $c$ and $b$.
The plot is made for the $z$-linear  model.
}
\end{figure}

\begin{center}
{\em\Large 6. Summary}
\end{center}

We have presented the approximate, analytical formulation of
the reionized CMB spectra $C^{XX}_l$ generated by RGWs.
Even though its approximate nature implies its application
as a complement to the numerical codes,
it does improve our understanding CMB
and efficiently promote the analysis  of various effects
that reionization brings upon  $C^{XX}_l$.

The  reionization around $z\sim 11$ is
studied  by three simple homogeneous models,
i.e., a sudden reionization,
two extended reionizations
with ionization fraction $X_e(\eta)\propto \eta$
and $X_e(\eta)\propto z$.
The key parameter is $\kappa_r$,
the optical depth from the present back up to the start of reionization.
Given a value of $\kappa_r$ in each model,
the visibility function $V_r(\eta)$ follows,
which is approximately fitted by Gaussian type functions.
This procedure is similar to
the treatment of decoupling in our previous study.

Then the time integrations for polarization mode $\beta_l$
and temperature anisotropies mode $\alpha_l$ are carried out
approximately,
and  the resulting analytic expressions
consist of contributions by
RWGs  $h(\eta_d)$ and $\dot h(\eta_d)$ at the decoupling,
and by $h(\eta_r)$ and $\dot h(\eta_r)$ at the reionization
as well.
It is found that,
while  $h(\eta_d)$ and $\dot h(\eta_d)$
 generate the primary peaks  at $l\ge 100$,
$\dot h(\eta_r)$  produces bumps for $C_l^{EE}$ and $C_l^{BB}$ at $l\sim5$,
and   $h(\eta_r)$ enhances  $C_l^{TT}$ and $C_l^{TE}$ there.
The analytic $C_l^{XX}$ qualitatively agree with
those by the numerical computing, such as CAMB.

As a merit of our analytic approach,
the dependence of $C_l^{XX}$  upon
the optical depth $\kappa_r$ are explicitly given,
in terms of the coefficients $a_1(\kappa_r)$ and $a_2(\kappa_r)$
for the polarization $\beta_l(\eta_0)$,
and of the coefficients  $e^{-\kappa_r}$ and $(1-e^{-\kappa_r})$
for the temperature anisotropies $\alpha_l(\eta_0)$.
It is found that $a_1(\kappa_r)$ and $a_2(\kappa_r)$ vary with $\kappa_r$
more quickly than $e^{-\kappa_r}$ and  $(1-e^{-\kappa_r})$,
respectively.
Therefore, the polarization $\beta_l$ is more sensitive to $\kappa_r$
than the temperature anisotropies $\alpha_l$ does.
A larger $\kappa_r$ gives higher  $a_2(\kappa_r)$
and lower  $a_1(\kappa_r)$,
i.e., yielding  higher bumps and lower primary peaks
in $C_l^{EE}$ and $C_l^{BB}$.
Thus there is a degeneracy of $\kappa_r$
with the normalization of the initial amplitude $A$ of RGWs.
Besides, $\kappa_r$ also has a  weak
degeneracy with the spectral index $\beta_{inf}$ of RGWs
since a larger $\beta_{inf}$
enhances the primary peaks on small scales.
The analytical  $C_l^{EE}$ and $C_l^{BB}$
also suggest possible ways to break these two kinds of degeneracies.

Besides $\kappa_r$, our formulation also demonstrates
the effects of the reionization time $\eta_r$
and the reionization duration $\Delta\eta_r$.
For a fixed $\kappa_r$,
the height of bump is proportional to $\Delta\eta_r$,
and the location $l$ of  bump depends on $\eta_r$
in such a way $l\sim k(\eta_0-\eta_r)$
that a later reionization (larger  $\eta_r$) yields
a bump at larger angular scales (smaller $l$).

Given a fixed set of parameters $\kappa_r$, $\eta_r$, and
$\Delta\eta_r$, the $\eta$-linear and $z$-linear models yield
similar bumps in $C_l^{EE}$ and  $C_l^{BB}$ . Thus our analytical
formulation is unable to rediscover the reionization history
$X_e(\eta)$ from $C_l^{XX}$.

These analytical results tell that
studies of reionization
by means of  CMB temperature anisotropies and polarization
not only requires  sufficient observational data,
but also need detailed studies of the reionization process itself
and more realistic modeling.

The reionization process also significantly affects
the possible detections of RGWs via the observations of $C_l^{XX}$.
In  particular,
it is found that
the reionization causes a shift of the zero multipole $l_0$ of
the cross spectrum $C_l^{TE}$
by a substantial amount $\Delta l\sim 20$,
which is also $\kappa_r$-dependent.
The effect of reionization need be properly included,
before one can apply the zero multipole method
to extract the traces of RGWs from the observed $C_l^{TE}$.

\

ACKNOWLEDGMENT: T.Y Xia's work has been partially
supported by Graduate Student Research Funding from USTC.
Y.Zhang's research work has been supported by the CNSF
No.10773009, SRFDP, and CAS.
We thank Dr. Zhao for interesting discussions.


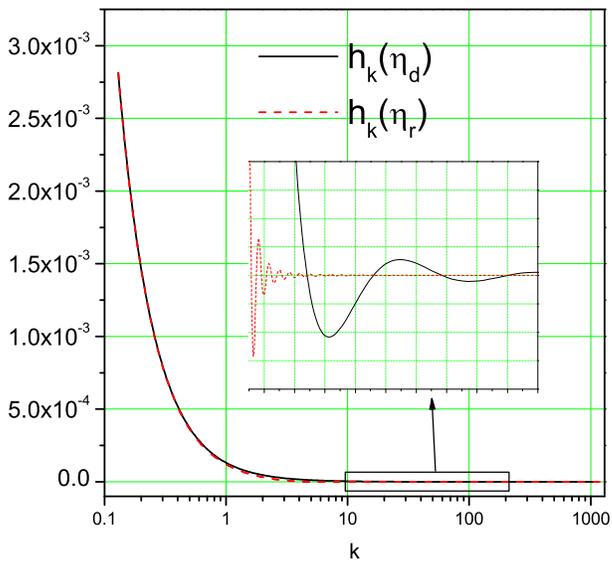
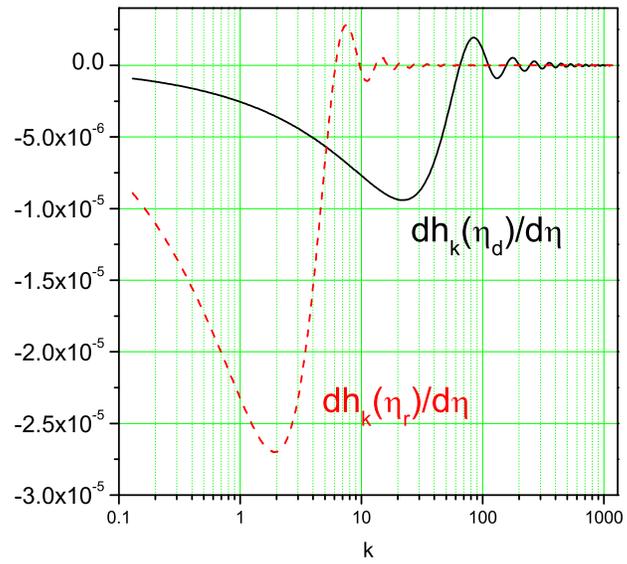
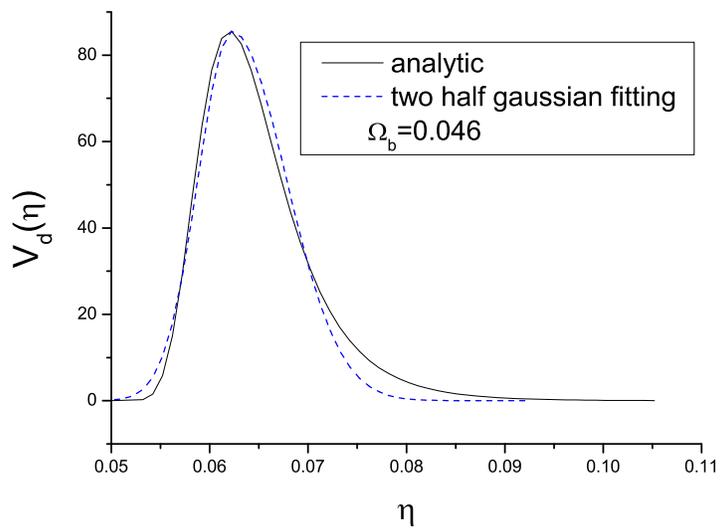
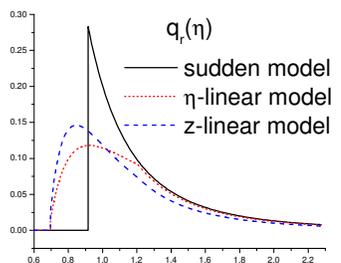
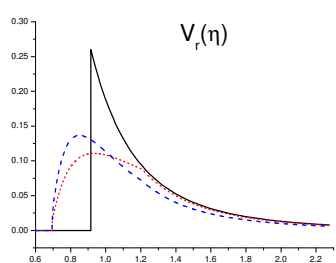
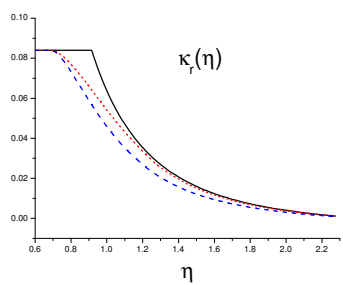
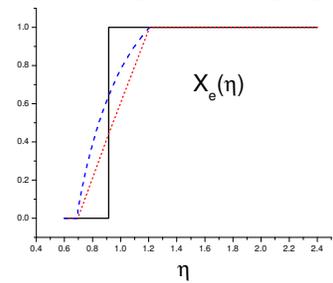
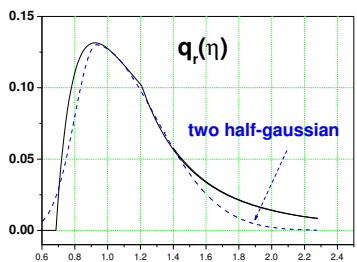
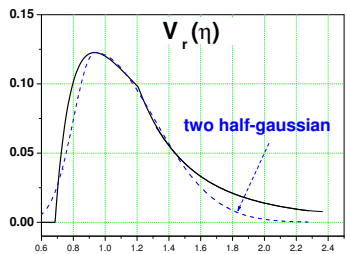
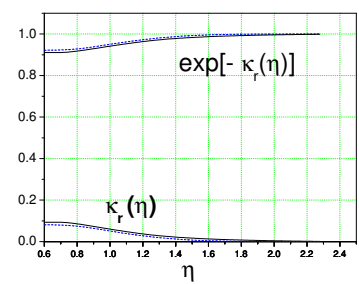
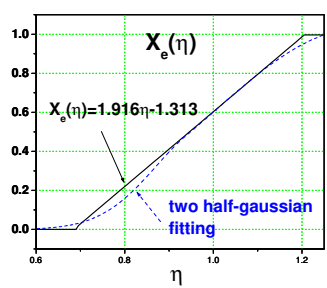
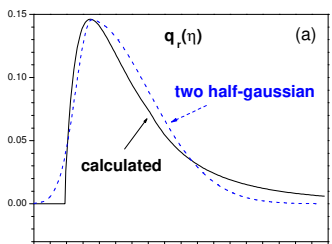
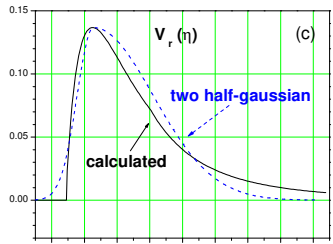
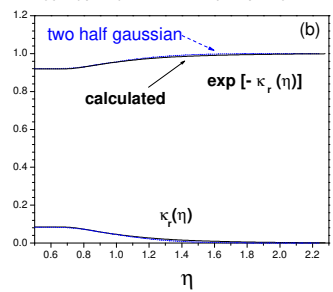
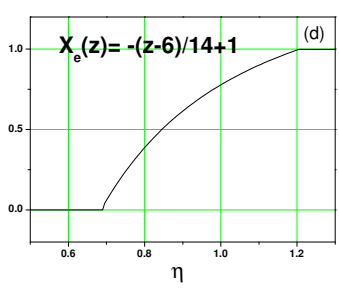
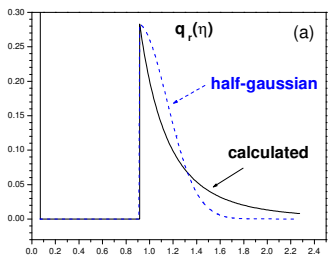
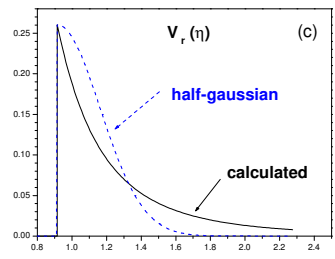
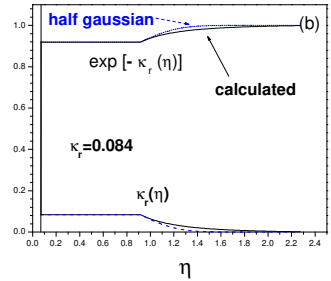
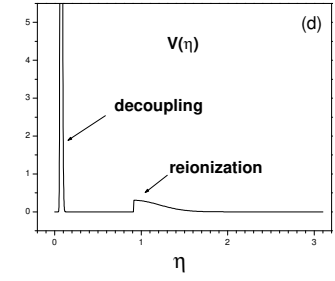
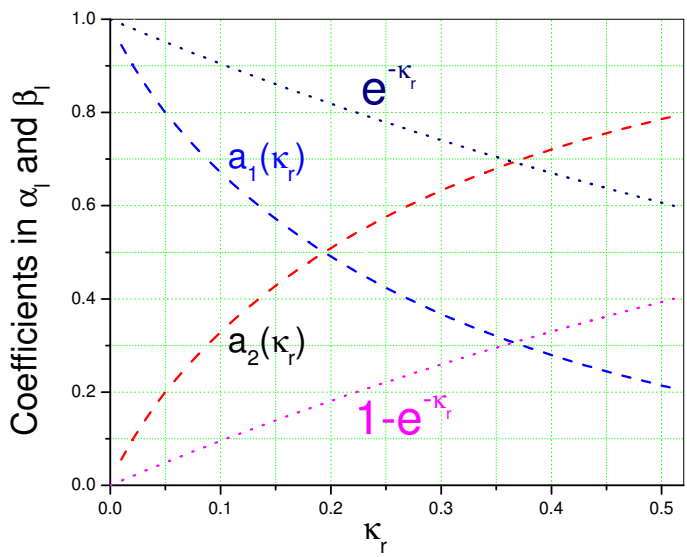
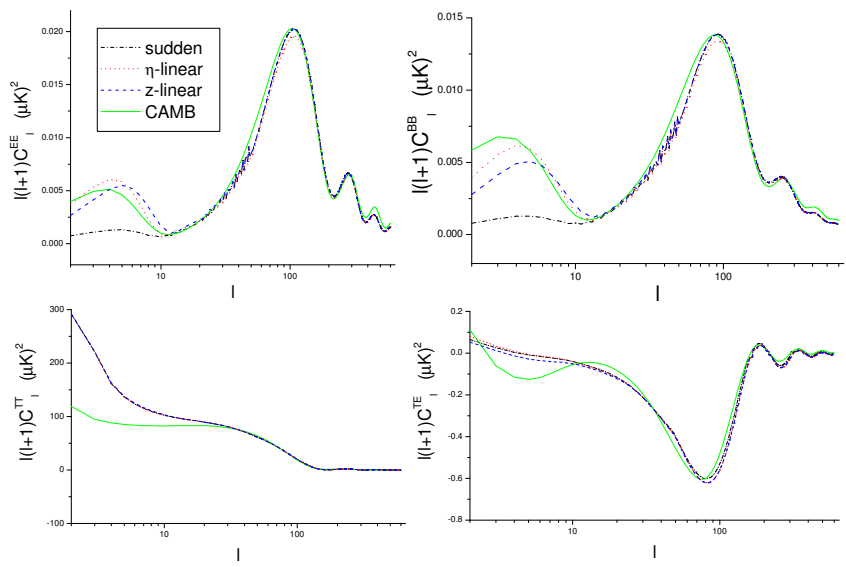
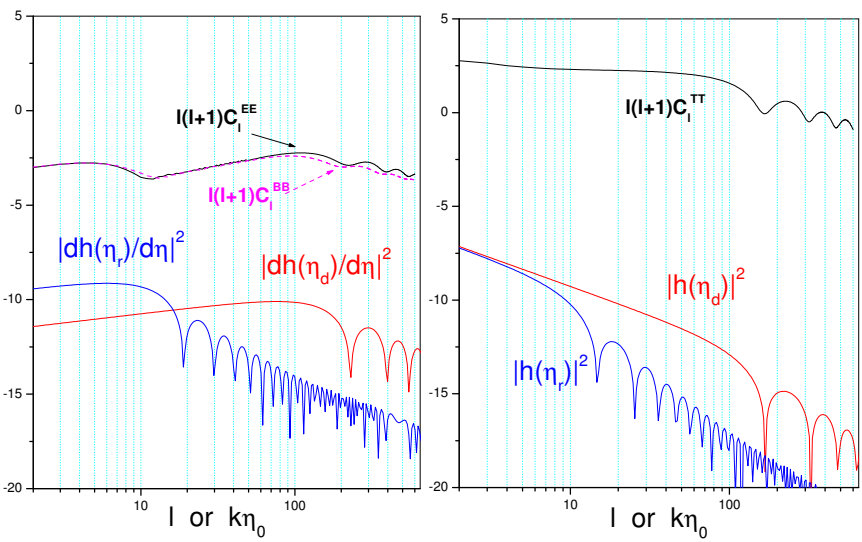
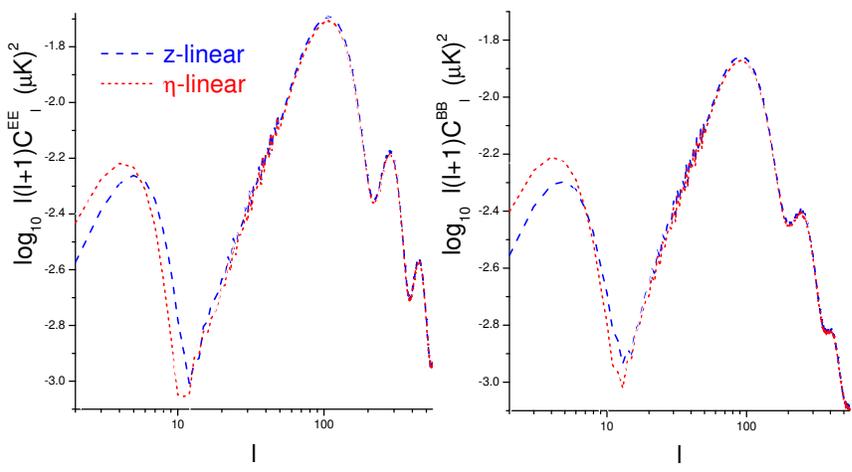
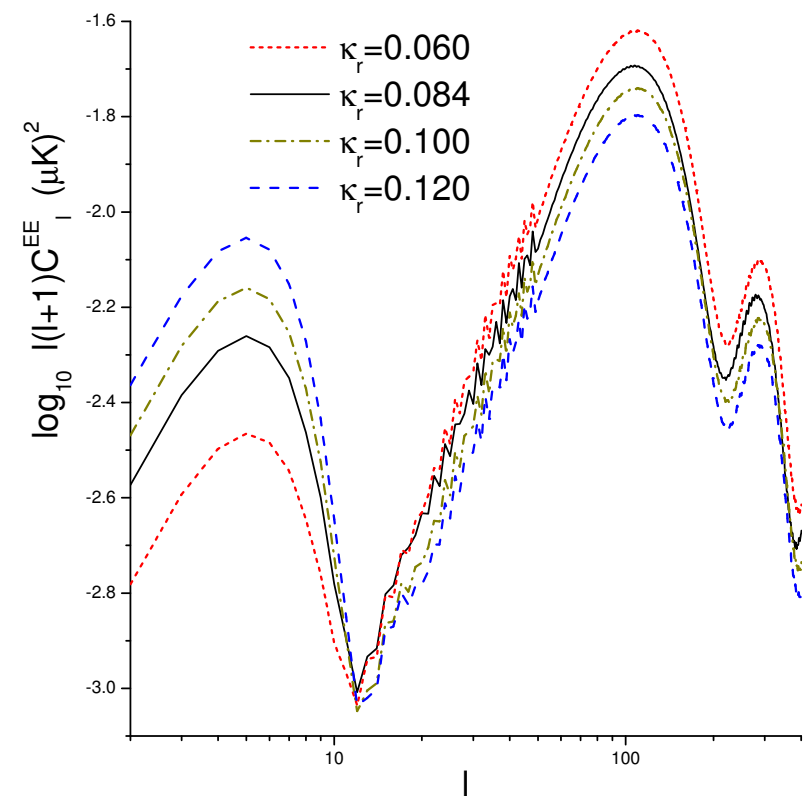
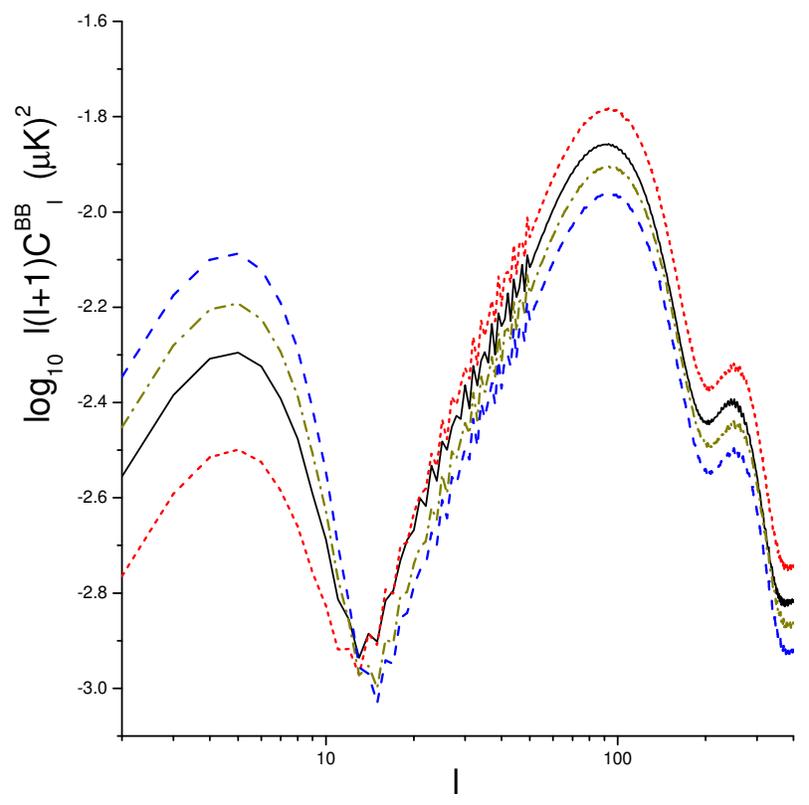
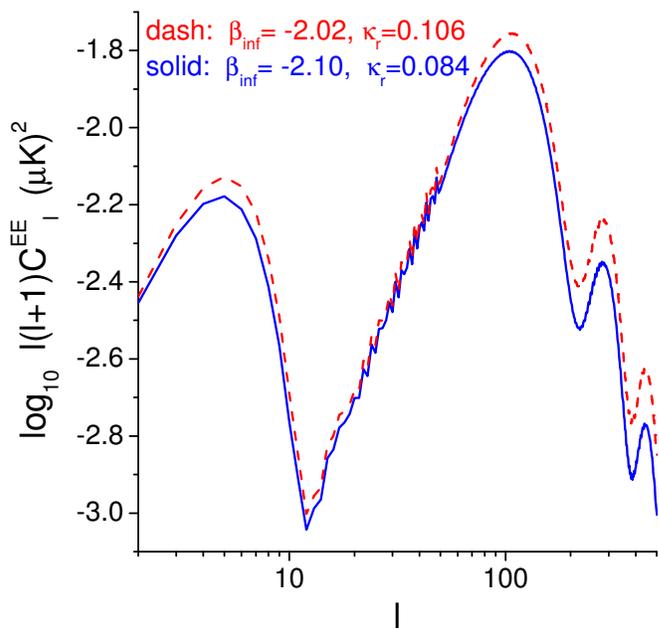
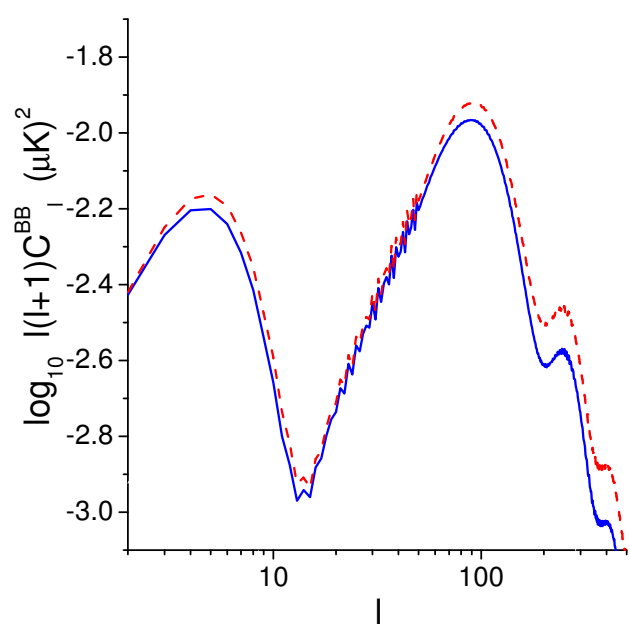
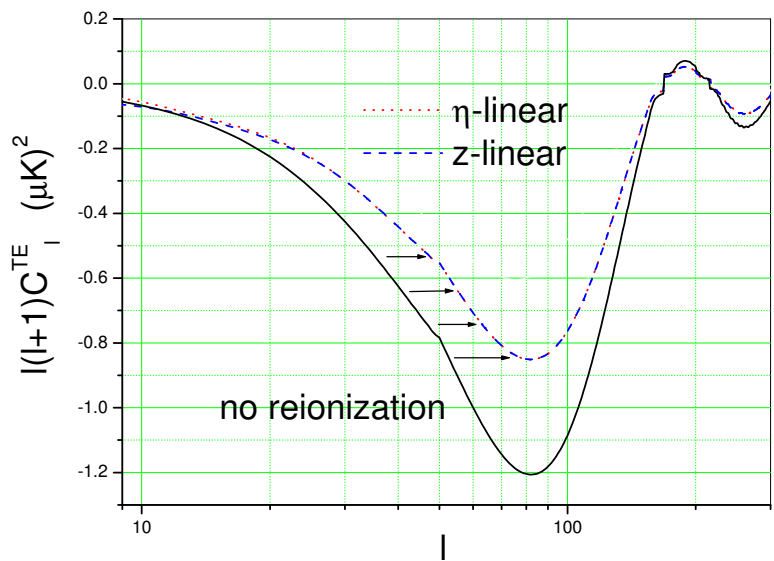
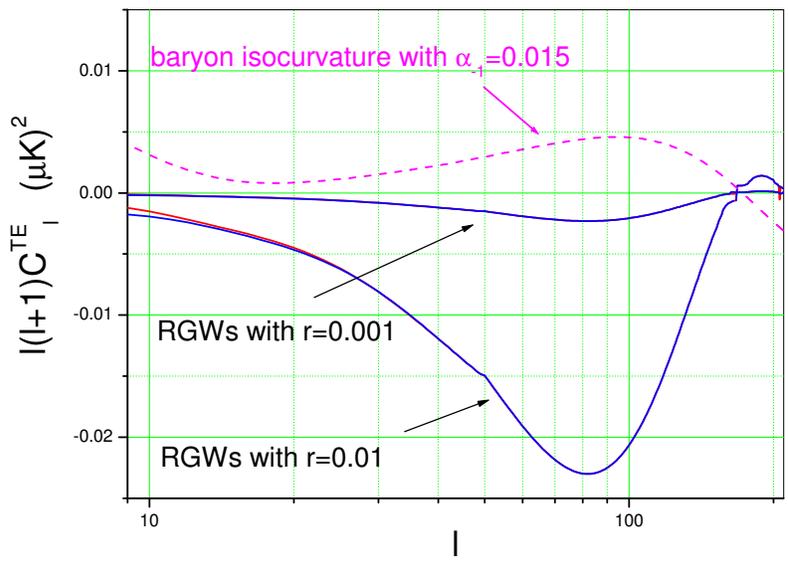
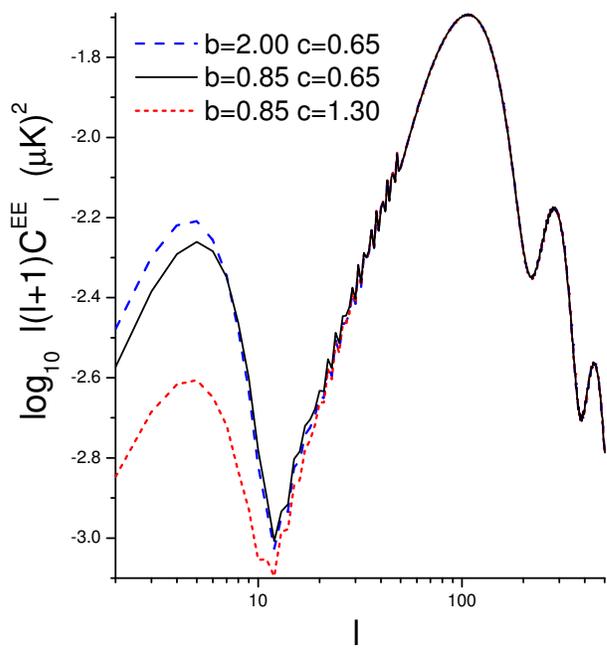
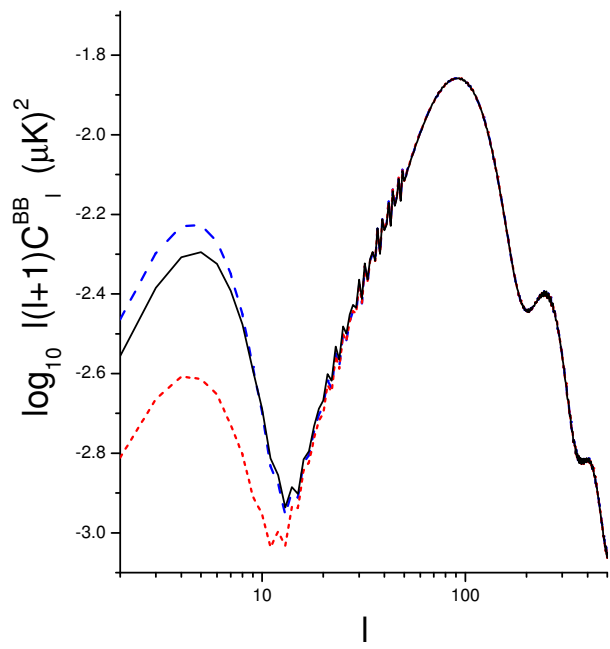
\end{document}